\documentclass[useAMS,usenatbib
]{mnras}
\usepackage{natbib}
\usepackage{graphicx}
\usepackage[section]{placeins}
 
\usepackage{yfonts}
\usepackage{times} 
\usepackage{amsmath}
\usepackage{amssymb}
\usepackage{tikz}
\usepackage{xcolor}

\def\icarus{Icarus}

\def\aj{Astron.\ J. }

\def\apjl{Astrophys.\ J. }
\def\apjs{Astrophys.\ J.\ (Supp.) }
\def\aap{Astron.\ Astrophys. }

\def\bssA_{Bull.\ Seismol.\ Soc.\ Am. }

\def\gcA_{Geochim.\ Cosmochim.\ Acta }

\def\mnras{Mon.\ Not.\ R.\ Astron.\ Soc. }

\def\prA_{Phys.\ Rev.\ A }

\defcitealias{NamouniMorais17b}{Paper I}

\begin{document}

\title[Disturbing function for arbitrary inclinations]{The disturbing function for asteroids with arbitrary inclinations}
\author[F. Namouni and M. H. M. Morais]{F. Namouni$^{1}$\thanks{E-mail:
namouni@obs-nice.fr (FN) ; helena.morais@rc.unesp.br (MHMM)} and  M. H. M. Morais
$^{2}$\footnotemark[1]\\
$^{1}$Universit\'e C\^ote d'Azur, CNRS, Observatoire de la C\^ote d'Azur, CS 34229, 06304 Nice, France\\
$^{2}$Instituto de Geoci\^encias e Ci\^encias Exatas, Universidade Estadual Paulista (UNESP), Av. 24-A, 1515 13506-900 Rio Claro, SP, Brazil}

\date{Accepted 2017 October 06 . Received 2017 October 06 ; in original form 2017 July 17}

\maketitle

\begin{abstract}
The  classical disturbing function of the three-body problem widely used in planetary dynamics studies is  an expansion of the gravitational interaction of the three-body problem with respect to zero eccentricity and zero inclination. This restricts its validity to nearly coplanar orbits. Motivated by the dynamical study of asteroids, Centaurs and transneptunian objects with arbitrary inclinations, we derive a series expansion of the gravitational interaction with respect to an arbitrary reference inclination that generalises our  work on the polar and retrograde disturbing functions. The new disturbing function, like the polar one, may model any resonance as expansion order is unrelated to resonance order. The powers of eccentricity and inclination of the force amplitude of a $p$:$q$ resonance depend only on the parity of the resonance order $|p-q|$. Disturbing functions with non zero reference inclinations are thus physically different from the classical disturbing function as the former are based on the three-dimensional three-body problem and the latter on the two-dimensional one. We illustrate the use of the new disturbing function by showing that what is known as pure eccentricity resonances are intrinsically dependent on inclination contrary to the prediction of the classical disturbing function. We determine the inclination dependence of the resonance widths of the 2:1 and 3:1 prograde and retrograde inner resonances with Jupiter as well as those of the asymmetric librations of the 1:2 and 1:3 prograde outer resonances with Neptune.
\end{abstract}


\section{Introduction}\label{section1}
The recent discoveries that Centaurs and transneptunian objects (TNOs) with  polar and retrograde inclinations may be captured in mean motion resonance by the solar system's planets \citep{MoraisNamouni13b,MoraisNamouni17b,Wiegert17} as well as the increasing number of TNOs on polar orbits  \citep{Gladmanetal09,Chen16} opened the new field of study on the origins and evolution of {asteroids} with arbitrary inclinations.  Hopes for an analytical understanding of the intricacies of asteroid dynamics  and capture at arbitrary inclinations have been hampered by the inadequacy of the classical analytical tools. Chief among them is the most basic one: the disturbing function of the three-body problem, the series expansion in the vicinity of circular {and coplanar} orbits of the gravitational interaction of two-bodies that revolve around the sun. Lacking such a basic tool for arbitrary inclinations, analytical studies could concentrate only on nearly coplanar retrograde orbits  \citep{MoraisNamouni13a} whereas intensive numerical simulations probed capture mechanisms and dynamical stability at arbitrary inclination \citep{NamouniMorais15,MoraisNamouni16,NamouniMorais17c}.  Of the most interesting and unexpected numerical results is the high  efficiency of resonance capture for polar and retrograde inclinations.  The basic physical origin of such  efficiency is known to be the encounter geometry as the retrograde asteroid encounters the planet for a shorter duration and larger velocity than if it were on a prograde orbit making retrograde resonances harder to destabilise \citep{NamouniMorais17}. A more precise and quantitative analytical  understanding of how capture operates would be possible if a disturbing function were derived to cover all inclinations and not only nearly coplanar prograde motion  \citep{EllisMurray00} and nearly coplanar retrograde motion \citep{MoraisNamouni13a}.

A step in the right direction was recently achieved with the disturbing function for polar Centaurs and TNOs \citep{NamouniMorais17b} --hereafter \citetalias{NamouniMorais17b}. A series expansion was carried out literally for nearly  polar motion in powers of eccentricity and inclination cosine. The resulting disturbing function was quite different from the classical one {for} nearly circular coplanar prograde motion \citep{EllisMurray00} and challenged long held beliefs on the dependence of the force amplitude of resonant terms on the asteroids's orbital elements.  For instance, it was found that any $p$:$q$ mean motion resonance has,  to lowest order in eccentricity, a force amplitude proportional to eccentricity $e$ if the resonance order $|p-q|$ is odd and proportional  to $e^2$ if the order is even. Such behaviour originates from the 
fact that the expansion is carried out with respect to a three-dimensional unperturbed configuration of reference inclination $I_r=90^\circ$ unlike the classical expansion that is built on the two-dimensional three-body problem of reference inclination $I_r=0^\circ$. This difference adds a new degree of freedom to the disturbing function embodied by new two-dimensional Laplace coefficients that bear the semimajor axis dependence of the force amplitude.\footnote{These aspects are discussed in detail in Section \ref{section4}.} The polar disturbing function helped identify TNO (471325) as the first transneptunian object in polar resonance with Neptune \citep{MoraisNamouni17b}. 

The three available disturbing functions: the classical one for nearly circular coplanar prograde motion  \citep{EllisMurray00},  its companion, the nearly circular coplanar retrograde expansion  \citep{MoraisNamouni13a} and the polar expansion seem to cover large portions of inclination space for the study of resonance dynamics. However, this situation is not satisfactory as it poses the problem of in-between inclinations. If an asteroid's orbit has an inclination of $45^\circ$ or its polar symmetric $135^\circ$ which one of the three expansions must be used? It was shown in \citetalias{NamouniMorais17b} that below an inclination of $55^\circ$ (and beyond the polar symmetric $125^\circ$) the polar disturbing function to fourth order in eccentricity and inclination cosine is imprecise and {higher} expansion orders are required. The classical disturbing function and its retrograde companion suffer the same limitations at different inclination thresholds \citep{Knezevic95}. Therefore the range of intermediate inclinations is substantial and may not be modelled by any of the known disturbing functions.  {In order to produce  an analytical theory of mean motion resonance capture and crossing at arbitrary inclination,} the opportune action is to develop an expansion with respect to eccentricity and an arbitrary and unspecified reference inclination $I_r$. If this is possible, then even for inclinations such as $10^\circ$ or $100^\circ$, the corresponding disturbing function with an equal reference inclination will be far more accurate than the classical expansion or the polar one even if they can be used, and will require fewer terms in the series to model their dynamics. What is said more precisely in mathematical terms is that approximating a function $f(x)$ by Taylor-expanding it  in the vicinity of $x_0=0$ in order to apply it to the vicinity of an $x_1\gg0$ is neither precise nor acceptable when  $f(x)$ can be Taylor-expanded in the vicinity of $x_1$ directly.  

In this {paper}, we expand the gravitational potential of an asteroid {with an eccentric and inclined orbit, that interacts with a planet on a circular orbit, around zero eccentricity and a reference} orbital plane of unspecified inclination $I_r$.  As noted in \citetalias{NamouniMorais17b}, disturbing functions come in two types: the first is that of expansions with respect to eccentricity and some function of inclination. Detailed expositions and historical perspectives on the classical disturbing function for nearly coplanar orbits are found in \cite{BC61}. The polar series expansion of \citetalias{NamouniMorais17b} is another such example. The second type is that of expansions with respect to the semi-major axis ratio that were recently reformulated in modern language by \cite{LAskar10}. The first type is the most widely used in nearly all subjects in planetary dynamics because of a significant shortcoming in the second type. Indeed, series expansions with respect to semi-major axis ratio $\alpha$ are notorious for their slow convergence if the condition $\alpha\ll 1$ is not {satisfied,} requiring an exceedingly large expansion order (see the discussion of the $\alpha$-expansion of the two-dimensional Laplace coefficients in Appendix \ref{app}). This feature confines the second type disturbing functions to the study of the perturbation of distant bodies such as the Sun on natural and artificial satellites. For our application needs, the dynamics of asteroids, Centaurs and TNOs that usually do not satisfy $\alpha\ll 1$, only series expansions of the first type are useful. 

We therefore develop a series expansion  in powers of eccentricity $e$ and inclination variable $s=\sin(I-I_r)$. Surprisingly the resulting disturbing function provided explicitly to fourth order in $e$ and $s$ is not particularly more involved than the polar expansion of \citetalias{NamouniMorais17b} itself a special case thereof for $I_r=90^\circ$.  Most of the new properties of the polar disturbing function are present in the general disturbing function and show that the latter two expansions belong to a class of disturbing functions based on the three-dimensional three-body problem unlike the classical disturbing function that is based on a perturbation of the two-dimensional problem through the assumption of nearly coplanar orbits for the asteroid and the planet. In Section \ref{section2}, we provide the literal expansion of the disturbing function with an arbitrary reference inclination and show how the new two-dimensional Laplace coefficients that are commonly functions of the semimajor axis ratio of the asteroid and the planet, become functions of the reference inclination $I_r$ as well. The properties of the new Laplace coefficients are discussed in Appendix \ref{app}. The reader who is not interested in the derivation details of the new disturbing function can skip Section \ref{section2} and find the expression of the series expansion in Section \ref{section3}. The properties of the new disturbing function and how it reduces to the classical one whose $I_r=0^\circ$, its companion expansion for retrograde motion whose $I_r=180^\circ$, and the polar expansion of \citetalias{NamouniMorais17b} whose $I_r=90^\circ$ are discussed in Section \ref{section4}. In Section \ref{section5}, we illustrate the importance of the new disturbing function by deriving the resonance width of pure eccentricity resonances whose resonant arguments do not include inclination through the absence of the longitude of ascending node in their expressions. According to the classical disturbing function these resonances are independent of inclination. We show that {this} conclusion is erroneous and pure eccentricity resonances depend strongly on inclination. In order to perform the resonance {width} calculations and apply them to prograde and retrograde inner resonances with Jupiter and outer resonances with Neptune, we develop a new pendulum model of resonance based on two resonant harmonics that is able to reproduce the known asymmetric librations of outer resonances \citep{Beauge94,Bruno,Malhotra96,WinterMurray97}. Section \ref{section6} contains our concluding remarks. 

\section{Literal expansion with arbitrary reference inclination}\label{section2}
We perform the literal expansion for arbitrary inclination using the same notations and steps of the literal expansions for nearly polar orbits (\citetalias{NamouniMorais17b}). The notations are also consistent with the development of the literal expansions for  nearly coplanar prograde orbits of \citep{ssdbook} and nearly coplanar retrograde orbits \citep{MoraisNamouni13a}.

We consider an asteroid that moves under the gravitational influence of the sun  of mass $M_{\star}$ and a planet of mass $m^\prime\ll M_{\star}$. The planet's {orbit} is  circular of  radius $a^\prime$ and longitude angle $\lambda^\prime$.  The reference plane is defined by the sun-planet orbit. The asteroid's osculating Keplerian orbit with respect to the star has semi-major axis $a$, eccentricity $e$, inclination $I$, true anomaly $f$, argument of pericentre $\omega$, and  longitude of ascending node $\Omega$.  After normalising all distances to $a^\prime$, the  disturbing function  reads:
\begin{equation}
R= {G\,m^\prime}{a^{\prime-1}} (\Delta^{-1}-r\,\cos{\psi})\equiv {G\,m^\prime}{a^{\prime-1}}\bar{R},
\end{equation}
where $r=\alpha (1-e^2)/(1+e \cos f)$ is the orbital radius of the asteroid and $\alpha=a/a^\prime$ is the asteroid's normalised semimajor axis that may be larger or smaller than unity,  $\psi$ is the angle between the radius vectors of  the planet and the asteroid,
$\Delta^2=1+r^2-2\,r\,\cos{\psi}$ is the planet-asteroid relative distance and
\begin{equation}
\cos\psi=\cos(\Omega-\lambda^\prime)\cos(f+\omega )-\sin(\Omega-\lambda^\prime)\sin(f+\omega )\cos I.  \label{cospsi}
\end{equation}
The first term of $\bar{R}$  is  the direct perturbation that we denote $\bar{R}_d$ and  the second term, that we denote $\bar{R}_i$, is the indirect perturbation that comes from the reflex motion of the star under the influence of the planet as the standard coordinate system is chosen to be  centred on the star.

The classical disturbing function is expanded in powers of $e$ and $\sin^2(I/2)$ with respect to a reference inclination $I_r=0^\circ$. It is therefore valid for nearly coplanar prograde motion as $\sin^2(I/2)$ vanishes for $I=I_r$.  The classical disturbing function may be transformed into an expansion in terms of $e$ and $\cos^2(I/2)$ valid for nearly coplanar retrograde orbits as $\cos^2(I/2)$  vanishes for the reference inclination $I_r=180^\circ$ \citep{MoraisNamouni13a}. For nearly  polar orbits and hence $I_r=90^\circ$, the disturbing function may be expanded in powers of $e$ and $\cos I$ (\citetalias{NamouniMorais17b}). For an arbitrary reference inclination $I_r$, we write $I=I_r+\delta I$ in the expression of  $\cos\psi$ and reduce it as follows:
\begin{eqnarray}
\cos\psi&=&\cos(\Omega-\lambda^\prime)\cos(f+\omega )\nonumber \\
&&-\sin(\Omega-\lambda^\prime)\sin(f+\omega )\cos I_r +\Psi\\
\Psi&=&\left([1 - (1 - s^2)^{1/2} ] \cos I_r  +s \sin I_r \right) \times \nonumber \\
&&\sin(f+\omega )\sin(\Omega-\lambda^\prime)
\end{eqnarray}
where $s=\sin \delta I= \sin (I-I_r)$. At this point, we note that the interpretation of the new disturbing function is twofold because of the presence of the reference inclination. We can choose and expand the gravitational interaction with respect to $e$ and $s$ and obtain a series that generalises the classical nearly coplanar and nearly polar disturbing functions. This will be done shortly. The second interpretation arises if we set $s\equiv0$  at this level of the expansion development or in the full series expansion given in Section \ref{section3}, then the new disturbing function is only an expansion with respect to eccentricity and keeps a full (unexpanded) dependence on inclination as the reference inclination is actually the asteroid's inclination for $s\equiv0$. This aspect will be discussed further in Section \ref{ss4.3}.   
With this in mind, we may write 
the asteroid-planet relative distance as:
\begin{eqnarray}\Delta^2&=&1+r^2-2 r
\cos(\Omega-\lambda^\prime)\cos(f+\omega )\nonumber \\
&&+2r\sin(\Omega-\lambda^\prime)\sin(f+\omega )\cos I_r 
-2r\Psi
\end{eqnarray} 
Expanding the direct perturbation term $\Delta^{-1}$ in the vicinity of $\Psi=0$,  we may write:
\begin{eqnarray}
\Delta^{-1} &=& \sum_{i=0}^{\infty} \frac{(2 i) !}{2^i(i !)^2} (r\Psi)^{i}  {\Delta_{0}^{-(2\,i+1)}},\label{Deltaexp}
\end{eqnarray}
where $\Delta_0^2=1+r^2-2\,r\,\cos(\Omega-\lambda^\prime)\cos(f+\omega )+2r\sin(\Omega-\lambda^\prime)\sin(f+\omega )\cos I_r$. Defining  $\epsilon=r/\alpha-1=\mathcal{O}(e)$ and expanding $\Delta_{0}^{-(2\,i+1)}$ around $\epsilon=0$, {we get:}
\begin{equation}
{\Delta_{0}^{-(2i+1)}} =
\left( 1+\sum_{l=1}^{\infty}\frac{\epsilon^l\,\alpha^l}{l!}\,\,\frac{d^l}{d \alpha^l} \right) {\rho^{-(2i+1)}}, \label{Delta0}
\end{equation}
\begin{eqnarray}
{\rho^{-(2i+1)}}&=&   [1+\alpha^2-2\,\alpha\,\cos(\Omega-\lambda^\prime)\cos(f+\omega )+\nonumber \\&&+2\alpha\sin(\Omega-\lambda^\prime)\sin(f+\omega )\cos I_r]^{-(i+1/2)}. \label{rho2i}
\end{eqnarray}
The next step is to develop the function $\rho^{-(2i+1)}$ into a two-dimensional Fourier series with respect to the angles $f+\omega$ and $\Omega-\lambda^\prime$ as follows:
\begin{equation}
 {\rho^{-(2i+1)}}=\! \! \! \! \! \! \! \! \! \!  \sum_{
{\scriptsize
\begin{array}{c}
-\infty<j,k<\infty\\ 
j+k \ {\rm even}
\end{array}}}\! \! \! \frac{1}{4}b_{i+1/2}^{jk}(\alpha,I_r) \cos[k(f+\omega )+j(\Omega-\lambda^\prime)], \label{RHO}
\end{equation}
\begin{eqnarray}
&&b_{s}^{jk}(\alpha,I_r) = \frac{1}{\pi^2}\int_0^{2\pi}  \int_0^{2\pi} {\cos(j u+kv) \ du\, dv} \times\label{Lap0}  \\
& &\ \ \ \ \ \ \ \ \ \ \ \ {[1+\alpha^2-2\alpha (\cos u \cos v-\sin u\sin v \cos I_r)]^{-s}},\nonumber 
\end{eqnarray}
where $b_{s}^{jk}(\alpha,I_r)$ are two-dimensional Laplace coefficients that reduce to those of the polar expansion for $I_r=90^\circ$. The index $s$ is a half integer and should not be confused with the inclination variable. We chose these notation conventions to be in accord with the significant body of literature on the classical disturbing function both historical and contemporary. The series (\ref{RHO}) is summed over even $j+k$ owing to the invariance of the function $\rho^{2i+1}$ (\ref{rho2i}) with respect to the variable change $(f+\omega+\pi,\Omega-\lambda^\prime+\pi)$ that makes $b_{s}^{jk}=0$ if $j+k$ is odd. The appearance of the two-dimensional Laplace coefficients is related to the presence of the two independent angles $f+\omega$ and $\Omega-\lambda^\prime$ in the expression of $ {\rho^{-(2i+1)}}$ just like the case of the polar expansion (\citetalias{NamouniMorais17b}) but unlike to the expansions of nearly coplanar orbits where those angles enter only as the sum  $f+\omega\pm(\Omega-\lambda^\prime)$ where the $\pm$ signs are for prograde and retrograde orbits respectively \citep{MoraisNamouni13a}. The properties of the two-dimensional Laplace coefficients with {an} arbitrary reference inclination are discussed in Appendix \ref{app}.  
 
 Substituting the series (\ref{RHO}) into the expression (\ref{Delta0}) and the latter into the expansion (\ref{Deltaexp}), the direct part of the perturbation is  written as the following series: 
\begin{equation}
\bar{R}_d =\!\! \!\!\!\!\!\!\!\!\!\!\!\!\!\sum_{
{\scriptsize\begin{array}{c}
0\leq i,l<\infty\\ 
-\infty<j,k<\infty\\ 
j+k \ {\rm even}
\end{array}}}\!\!\!\!\!\!\!\!\!\!\!\! \!\!\!\frac{1}{l!} \frac{(2 i) !}{2^{i+2}(i !)^2} (  r \Psi )^{i} \epsilon^l A_{i,j,k,l}\cos[k(f+\omega )+j(\Omega-\lambda^\prime)],\label{RX}
\end{equation}
where 
\begin{equation}
A_{i,j,k,l}=\alpha^l D^l  b_{i+1/2}^{jk}, \label{Aijkl}
\end{equation}
satisfy the same restriction on the two-dimensional Laplace coefficients in that $A_{i,j,k,l}=0$ if $j+k$ is {odd,} as well as additional  properties of the Laplace coefficients discussed in Appendix \ref{app}. Next, we express $\Psi$ and $r$ in terms of   the mean {longitude} and the longitude of pericentre using {the classical elliptic expansions}:
\begin{eqnarray}
\sin f&=& 2 (1-e^2)^\frac{1}{2}\sum_{\sigma=1}^\infty \frac{d}{sde}J_\sigma(\sigma e)\sin [\sigma (\lambda-\varpi)],\nonumber\\
\cos f&=& -e +{2(1-e^2)}e^{-1}\sum_{\sigma=1}^\infty J_\sigma(\sigma e)\cos [\sigma (\lambda-\varpi)], \label{elliptic}\nonumber\\
\frac{r}{\alpha}&=& 1+\frac{e^2}{2}-2e\sum_{\sigma=1}^\infty \frac{d}{\sigma^2de}J_\sigma(\sigma e)\cos [\sigma (\lambda-\varpi)].
\end{eqnarray}

Upon substituting the expressions (\ref{elliptic}) into the direct part of the perturbation (\ref{RX}) and truncating it to order $N$ in eccentricity $e$ and inclination $s$, we obtain a series that needs to be transformed further to model a general $p$:$q$ resonance. This classical reduction step for literal expansions arises because seemingly unrelated arguments actually pertain to the same $p$:$q$ resonance. For example, among the various terms that appear to second order in $s=\sin\delta I$ and zero order in eccentricity, there are: 
\begin{eqnarray}
T_1&=&\frac{\alpha^2s^2}{64}  A_{2,j,k,0} \cos [k \lambda-j \lambda^\prime+(j-k) \Omega ], \\
   T_2&=&-\frac{3\alpha^2s^2\cos (2I_r)}{512} A_{2,j,k,0}  \times \\&&\ \ \ \ \ \ \ \cos [(k-2)\lambda-(j+2) \lambda^\prime+(j-k+4)   \Omega]. \nonumber
\end{eqnarray}
{As according to equation (\ref{RX}), the indices $j$ and $k$  can take any positive and negative integer values independently of each other}, both these terms may be made to correspond to the same resonance $p$:$q$ by choosing $j=p$, $k=q$ for $T_1$ and $j=p-2$ and $k=q+2$ for $T_2$ {to find}:
\begin{eqnarray}
T_1&=&\frac{\alpha^2s^2}{64}  A_{2,p,q,0} \cos [q \lambda-p \lambda^\prime+(p-q) \Omega ], \\
   T_2&=&-\frac{3\alpha^2s^2\cos (2I_r)}{512} A_{2,p-2,q+2,0}  \times \\&&\ \ \ \ \ \ \ \cos [q\lambda-p \lambda^\prime+(p-q)   \Omega]. \nonumber
\end{eqnarray}
Furthermore for $p\neq0$ and $q\neq 0$, $(-p)$:$(-q)$  cosine terms obviously correspond  to the $p$:$q$ resonance  as the series (\ref{RX}) is summed over positive as well as negative $k$ and $j$ {as noted previously}. Changing $p$ and $q$ to their opposites in $T_1$ and $T_2$ produces two new terms, $T_3$ and $T_4$, that correspond to the $p$:$q$ {resonance}:
\begin{eqnarray}
T_3&=&\frac{\alpha^2s^2}{64}  A_{2,p,q,0} \cos [q \lambda-p \lambda^\prime+(p-q) \Omega ], \\
   T_4&=&-\frac{3\alpha^2s^2\cos (2I_r)}{512} A_{2,p+2,q-2,0}  \times \\&&\ \ \ \ \ \ \ \cos [q\lambda-p \lambda^\prime+(p-q)   \Omega]. \nonumber
\end{eqnarray}
In the indices of $A_{0,p,q,0}$ and $A_{0,p+2,q-2,2}$, we use the properties  (\ref{ALap1} and \ref{ALap2}) of the two-dimensional Laplace coefficients. The secular terms may be obtained by setting $p=0$ and $q=0$ in $T_1$ and $T_2$ but not in $T_3$ and $T_4$ because the same term would be counted twice.\footnote{The pair $(0,0)$ is a fixed point of the transformation  ($p\rightarrow -p$, $q\rightarrow -q$).}

The indirect part of the disturbing function, $\bar{R}_i$ only requires the use of the elliptic expansions (\ref{elliptic}) to transform true anomalies into mean anomalies and effect the eccentricity expansion.  The resulting expressions of the direct, indirect and secular parts of the polar disturbing function are given in the next Section. 

\section{Disturbing function with arbitrary reference inclination}\label{section3}
For a $p$:$q$  cosine term and an expansion order $N$ in eccentricity $e$ and inclination $s$, the steps described in the previous Section show that the direct part of the disturbing function is given as:
\begin{eqnarray}
\bar{R}_d&=&\!\!\!\!\!\!\!\!\sum_{
{\scriptsize\begin{array}{c}
-N\leq k\leq N \\
|k|\leq m\leq N\\
0\leq n\leq N\\ 
m+n=N 
\end{array}}}\!\!\!\!\!\!\!\!
c^{k}_{mn}(p,q,\alpha,I_r)\, e^m s^n \ \cos \phi^{p:q}_k, \label{RY}\\
\phi^{p:q}_k&=&q \lambda -p \lambda^\prime +(p-q) \Omega-k\omega. \nonumber
\end{eqnarray}
The force coefficients $c^{k}_{mn}(p,q,\alpha,I_r)$ are given explicitly for the fourth order series $N=4$ and $k=0$, 1, 2, 3, and 4 in Tables \ref{t1} to \ref{t5}.  For negative $k$, the force coefficients may be obtained from the identity  $c^{-k}_{mn}(p,q,\alpha,I_r)=c^{k}_{mn}(-p,-q,\alpha,I_r+180^\circ)$ where $I_r+180^\circ$ is defined modulo $180^\circ$. Examination of the force coefficients shows the additional relationship:
 \begin{equation}
 c^{-k}_{mn}(p,q,\alpha,I_r)=c^{k}_{mn}(p,-q,\alpha,I_r+180^\circ). \label{cmkpk}
 \end{equation}
 
 {As no assumption was made regarding the  ratio $\alpha$ of the semimajor axis of the asteroid's orbit to that of the perturber's circular orbit, the force coefficients} are valid for $\alpha>1$ as well as $\alpha<1$ \citep{Williams69}. 
 
 {At this point, we caution the user of the arbitrary inclination disturbing function that the fourth order expansion provided in Appendix \ref{appB} does not model accurately the highest eccentricity orbits. However, it is a simple matter to code the literal disturbing function for arbitrary inclinations with the steps we provided in the previous Section using formal algebra software and easily obtain high eccentricity power terms in a relatively short runtime on a personal computer.}

The force coefficients $c^{k}_{mn}(p,q,\alpha,I_r)$ have an  important property related to the resonance order $o_r=|p-q|$.  Examination of  $c^{k}_{mn}(p,q,\alpha,I_r)$'s dependence on $A_{i,j,k,l}$ and recalling that $A_{i,j,k,l}=0$ when $j\pm k$ is odd  show that for an odd resonance order $o_r$, $c^{k}_{mn}(p,q,\alpha,I_r)=0$ when $k$ is even. Similarly for an even resonance order $o_r$, $c^{k}_{mn}(p,q,\alpha,I_r)=0$ when $k$ is odd. This property guarantees that the integer coefficient of the longitude of ascending node, $\Omega$, that reads $p-q+k$ is always even. 

These properties were encountered for the first time in the direct part of the disturbing function for nearly polar orbits (\citetalias{NamouniMorais17b}); they now appear as a general property of the disturbing function for arbitrary inclination and suggest that the classical disturbing function obtained with $I_r=0^\circ$ is a degenerate two-dimensional limit of the general expansion. More on this aspect in Section \ref{ss4.2}.

The arguments and force amplitudes up to and including fourth order of the disturbing function's indirect part for an arbitrary reference inclination $I_r$  are given in Table \ref{t6} to fourth order in eccentricity $e$ and inclination $s$. As in the polar expansion, the indirect part concerns only cosines of the type 1:$q$ with $0\leq q\leq 5$. For $q\neq 0$, arguments that include $q\lambda-\lambda^\prime$ may induce resonance whereas those that start out as $q\lambda+\lambda^\prime$ do not. 

The secular potential to order $N$ in eccentricity and inclination is obtained by setting $p=q=0$ in the literal expansion while taking care not to count the same terms twice in the expansion as explained in the previous Section. Its expression is found as:
\begin{equation}
\bar{R}_s=\frac{1}{2}b_\frac{1}{2}^{00}(\alpha, I_r)+\!\!\!\!\!\!\!\sum_{
{\scriptsize\begin{array}{c}
0\leq k,n\leq N \\
k\leq m\leq N\\
k,m  \ {\rm even}\\
m+n=N
\end{array}}} \!\!\!\!s^{k}_{mn}(\alpha,I_r) \, e^{m} s^n\ \cos (k\omega),\label{SecPot}
\end{equation}
where in contrast to that of the polar disturbing function of \citetalias{NamouniMorais17b}, the powers of inclination $s$ are no longer limited to even integers. This property is related to the symmetry of the expansion coefficients (and the Laplace coefficients) with respect to the polar plane and   is discussed further in Section \ref{section4} and Appendix \ref{app}. The {expressions} of the secular coefficients $s^{k}_{mn}(\alpha,I_r)$ are given in Table \ref{t7} for the $N=4$ series expansion. 

\section{Properties of the disturbing function with arbitrary reference inclination}\label{section4}
\subsection{Comparison with the disturbing function for nearly polar orbits}\label{ss4.1}
The polar disturbing function of \citetalias{NamouniMorais17b} ($I_r=90^\circ$) was the first disturbing function to be expanded with respect to a non-coplanar motion.\footnote{We consider only expansions in eccentricity and inclination but not semimajor axis  employed historically for the motion of planetary satellites. See  \cite{LAskar10} for a historical perspective and a modern presentation of such expansions.} It taught us that its properties differ significantly from those of the classical disturbing function whose $I_r=0^\circ$ \citep{EllisMurray00} and its companion retrograde expansion whose $I_r=180^\circ$ \citep{MoraisNamouni13a}. The main differences were three. First, the literal expansion of order $N$ was applicable to any resonance ratio $p$:$q$ in contrast to the nearly coplanar disturbing function that produces terms only for the ratios $p$:$p\pm N$ \citep{ssdbook}. For example the arguments $\phi^{3:10}_1=10 \lambda -3 \lambda^\prime -7\Omega-\omega$ and $\phi^{3:10}_3=10\lambda -3 \lambda^\prime -7\Omega-3\omega$ of the  3:10 resonance may be studied by the $N=4$ order series whereas with the classical disturbing function, in order to study the same arguments, the expansion order must be at least $N=7$. This is not a mere technical detail as the fact that the integers $p$ and $q$ are not related to one another through  $N$ in the polar expansion is rooted in the presence of the two independent angles $f+\omega$ and $\Omega-\lambda^\prime$ in the expression (\ref{RHO})  that in turn gave rise to the two-dimensional Laplace coefficients thus adding an extra degree of freedom to the problem. {As already stated in Section \ref{section2}}, these two angles show up in the classical disturbing function only as the combination $f+\omega\pm(\Omega-\lambda^\prime)$ where $\pm$ refers to the expansion with $I_r=0^\circ$ and $I_r=180^\circ$ respectively giving rise to the one-dimensional Laplace coefficients.  The independence of the two angles found in the polar expansion  is shared by the disturbing function with an arbitrary  reference inclination $I_r$  provided that  $I_r\neq 0^\circ$ and  $I_r\neq180^\circ$. {The new disturbing function is indeed applicable to any resonance ratio $p$:$q$. The expansion order $N$ limits only the integer coefficient $k$ of the argument of pericentre in the resonance angle  $\phi^{p:q}_k= q\lambda -p \lambda^\prime +(p-q)\Omega-k\omega$ to the values $|k|\leq N$ as can be seen in the expressions of the direct and secular parts of the disturbing function (\ref{RY},\ref{SecPot}).}

The next two properties found in the polar disturbing function are also shared by the new series expansion and originate in  the properties of the coefficients $c^{k}_{mn}(p,q,\alpha,I_r)$. One is  the fact that the powers of eccentricity $e$ and inclination $s$ in the force amplitudes of the new disturbing function are independent of the value of the resonance order and the other property is that only the parity of the resonance order decides the powers of $e$ and $s$ in the force amplitude of a $p$:$q$ resonance. As explained in \citetalias{NamouniMorais17b}, the {former} property does not {apply} for $I_r=0^\circ$ and $I_r=180^\circ$ since the powers of say the 1:4 resonance with $I_r=0^\circ$ are of the form $e^{|3-2m|}\sin^{2m}(I/2)$ to lowest order in eccentricity and inclination. If one seeks the force amplitude linear with respect to $e$ one has to {accept} the presence of the inclination factor  $\sin^{2}(I/2)$ in the force amplitude  because the sum of the powers of eccentricity and inclination must equal the resonance order $|p-q|=3$. With the new disturbing function, there is a term of power 1 in eccentricity and power zero in inclination $s$ of coefficient $c^{1}_{10}(1,4,\alpha,I_r)$ (Table \ref{t2}) as long as the chosen $I_r\neq 0^\circ$ and  $I_r\neq180^\circ$. We discuss {below} in more precise terms how the coefficients $c^{k}_{mn}(p,q,\alpha,I_r)$ behave when $I_r= 0^\circ$ and  $I_r=180^\circ$. 

The property regarding the force amplitude and the parity of resonance order tells us that all odd order $p$:$q$ resonances have force amplitudes proportional to eccentricity as $(c^{1}_{10}+c^{1}_{11}\, s)\, e$   to lowest order in eccentricity and inclination.   All even  order $p$:$q$ resonances have force amplitudes  of the form  $ c^{0}_{00}+ c^{0}_{01}\, s+ c^{2}_{20}\, e^2 + c^{2}_{02}\, s^2$  to lowest order in eccentricity and inclination. This universal binarity is  also a property of the new disturbing function with arbitrary inclination $I_r$  provided $I_r\neq 0^\circ$ and  $I_r\neq180^\circ$. The dependence on the smallest powers of $e$ and $s$ (linear and quadratic) is realised with the smallest possible integer $|k|$ in the argument $\phi^{p:q}_k= q\lambda -p \lambda^\prime +(p-q)\Omega-k\omega$. 

The polar and arbitrary inclination disturbing functions share the previous three general properties and even have some terms with identical expressions such as the pure eccentricity coefficients $c^k_{k0}$ that do not depend explicitly on $I_r$ but only implicitly through  the Laplace coefficients (\ref{Lap0}). The similarities however end there as the polar disturbing function is a special case of the general one for $I_r=90^\circ$. For instance, the two-dimensional Laplace coefficients for an arbitrary $I_r$ do not share all the symmetries of  their polar counterparts. Whereas for $I_r=90^\circ$, $b_s^{j(-k)}=b_s^{jk}$, the general Laplace coefficients only have reflection symmetry with respect to the polar plane as 
$b_s^{j(-k)}(I_r)=b_s^{jk}(I_r+180^\circ)$ modulo $180^\circ$ (see Appendix \ref{app}). One of the salient consequences of this symmetry is the presence of terms linear with respect to inclination $s$ in the secular potential (equation \ref{SecPot} and Table \ref{t7}) {that are not present in the polar disturbing function} ($I_r=90^\circ$). More generally, one can check easily that the coefficients of Tables \ref{t1} to \ref{t6} agree with those of \citetalias{NamouniMorais17b} for $I_r=90^\circ$.\footnote{Note that in this paper, the inclination variable $s=-\cos I$ for $I_r=90^\circ$, the opposite of that of \citetalias{NamouniMorais17b}. This means that all odd powers of $s$ will have an additional negative sign with respect to the coefficients derived in \citetalias{NamouniMorais17b}.}

The three properties of the polar expansion, generalised to the arbitrary inclination disturbing function signal different physical behaviours between  expansions performed with respect to the two-dimensional three-body problem, namely the classical disturbing function and its companion for retrograde orbits, and expansions performed with respect to the three-dimensional three-body problem such as the series we derive in this work.  

\subsection{Relationship to the classical and retrograde disturbing functions}\label{ss4.2}
To examine further the relationship between the nearly coplanar disturbing functions and the three-dimensional ones, we consider the disturbing function's coefficients for $I_r=0^\circ$ and $I_r=180^\circ$. We note that for $I_r=0^\circ$ or $180^\circ$, the inclination variable is $s=\pm \sin I$ where the $\pm$ sign is for the {classical} prograde and  retrograde disturbing functions respectively. However, the classical coplanar expansions are performed with respect to the variables $\sin(I/2)$ and $\cos(I/2)$.  This slight difference does not affect our comparison as consistency requires that we use the classical expansions near $I \sim 0^\circ$ and $180^\circ$ respectively, which in turn implies  that for the same cosine arguments, the classical disturbing function coefficients have extra $2^n$ factors for inclination terms with respect to those of the new disturbing {function} where $n$ is the inclination power in $s^n$.  An example of this situation will be given below using an inclination term of the $I_r=0^\circ$ disturbing function.

Inspection of Tables \ref{t1} to \ref{t7} shows that choosing $I_r=0^\circ$ or $180^\circ$ eliminates all terms with an odd power of the inclination $s$ in the disturbing function's direct part, indirect part as well as  the secular potential. This recovers the classical dependence of the nearly coplanar series on powers of $s^2$.  

In Appendix \ref{app}, we find that for $I_r=0^\circ$, the coefficients $A_{i,j,k,l}=0$ unless $j=k$ (\ref{AIr0}), {and} that for  $I_r=180^\circ$, the coefficients $A_{i,j,k,l}=0$ unless $j=-k$  (\ref{AIr180}). Applying these two properties to the terms $c^k_{k0}(p,q,
\alpha,I_r=0^\circ \ {\rm or }\ 180^\circ) \, e^{|k|}\, \cos \phi^{p:q}_k$ of Tables \ref{t1} to \ref{t6} (for which $k>0 $) reveals that the integer $q$ is no longer independent of $p$ but must satisfy $q-p =k$ for the prograde expansion and $q+p=k$ for the retrograde one.  For negative $k$, the force coefficients are obtained from equation (\ref{cmkpk}) as  $c^k_{k0}(p,-q,\alpha,I_r=180^\circ \ {\rm or }\ 0^\circ)$ for the prograde and retrograde expansions respectively,   implying that $-q+p=k$ for the prograde expansion (that now obeys the retrograde rule  $A_{i,j,k,l}=0$ unless $j=-k$)  and $-(q+p)=k$ for the retrograde  expansion. This leads to the general requirement $|p-q|=|k|$ for the prograde expansion and $|p+q|=|k|$ for the retrograde one. We thus recover the expansion order to resonance order relationship of the classical expansions  discussed at the beginning of this Section. 

{It is important to emphasise that this relationship between the integers $p$ and $q$  of a cosine term in the disturbing function is specific to the disturbing functions of nearly coplanar orbits ($I_r=0^\circ$ and $I_r =180^\circ$). There are no similar relationships for $I_r\neq 0^\circ$ and $I_r\neq180^\circ$. This indicates that if the general disturbing function is defined as a functional of the spatial dimension of the three-body problem that is chosen as a reference system with respect to which the series expansion is carried out, then the functional is extremal for dimension 2{,}  as the relationship $q=p\pm N$ removes a significant number of terms from the expansion --that is why we termed the coplanar disturbing functions degenerate. It is beyond the scope of this paper to provide the mathematical proof for this conjecture. }

In order to verify the agreement of the new disturbing function with the classical one, we recall  that the two-dimensional Laplace coefficients $b_s^{jk}(\alpha, I_r=
0^\circ)= 2 b_s^j(\alpha)$ where  $b_s^j(\alpha)$ is the classical (one dimensional) Laplace coefficient (\ref{Lap01}) (Appendix \ref{app}).  The prograde  force amplitudes {of eccentricity terms} discussed above are then reduced to those of the classical disturbing function of \cite{EllisMurray00} as follows: $e c^1_{10}(j-k,j, \alpha,I_r=0^\circ )=e f_{27}$ (term 4D1.1), $e^2 c^2_{20}(j-k,j, \alpha,I_r=0^\circ )=e^2 f_{45}$ (term 4D2.1),  $e^3 c^3_{30}(j-k,j, \alpha,I_r=0^\circ )=e^3 f_{82}$ (term 4D3.1), and  $e^4 c^4_{40}(j-k,j, \alpha,I_r=0^\circ )=e^4 f_{90}$ (term 4D4.1).   

The case of inclination terms is treated similarly with the difference that for a given inclination term, the resonances concerned are diverse. For instance, the term  $c^0_{02}(p,q,\alpha,I_r=0^\circ )\, s^2\,\cos \phi^{p:q}_0$ has a force coefficient $c^0_{02}(p,q,\alpha,I_r=0^\circ )={\alpha}  (A_{1, p-1, q+1, 0} +  A_{1, p+1, q-1, 0} -A_{1, p-1, q-1, 0} - A_{1, p+1, q+1, 0})/16$ and applies to one of  the three resonances $q=p$, $q=p-2$ and $q=p+ 2$ as the integers $j$ and $k$ in each of the $A_{1,j,k,0}$ are different. For the latter two resonances, there is only one term in the coefficient's expression that does not vanish. Explicitly, with $p=j+2$ and $q=j$, we have $s^2 c^0_{02}(j+2,j,\alpha,I_r=0^\circ )=\alpha s^2 A_{1,j-1,j-1,0}/16= \alpha s^2 b_{3/2}^{j-1}/8=  f_{57}s^2/4$ where the latter coefficient is that of the corresponding term 4D2.4 of \cite{EllisMurray00} --an additional  factor $1/4$ is present  because of the choice of  $s=\sin I$ instead of $\sin(I/2)$ as the inclination variable. More generally, it can be checked that the new fourth order disturbing function with $I_r=0^\circ$ agrees completely with the classical one \citep{EllisMurray00}.

\subsection{Two interpretations of the arbitrary inclination disturbing function}\label{ss4.3}
The last property of the new disturbing function that we discuss is its versatility. Unlike the classical disturbing function, its retrograde companion and the polar disturbing function, the new series expansion has two interpretations depending on whether the inclination variable $s$ is set to zero or kept as a free variable (Section \ref{section2}). In the latter case, the new disturbing function is a double series expansion with respect to eccentricity $e$ and inclination $s$. However {if} $s$ is identically set to zero, the new disturbing function becomes an expansion with respect to eccentricity $e$ only and remains unexpanded with respect to inclination. The identity $s\equiv 0$ then means that the reference inclination is just the asteroid's inclination. The dependence on inclination is transferred exclusively to  the new two-dimensional Laplace coefficients. 

The two interpretations of the new disturbing function have their advantages. For instance, when calculations are required to follow the time evolution of the asteroid's dynamics, the double $es$-expansion version is preferable as applying the Lagrange planetary equations to the $e$-expansion may prove difficult. Indeed, using the $e$-expansion with the Lagrange equations requires the evaluation of the Laplace coefficients and their derivatives with respect to inclination which in practice means the use of the $\alpha$-expansion of the Laplace coefficients derived in Appendix \ref{app}. To achieve good accuracy the $\alpha$-expansion in turn requires a large order (see  Appendix \ref{app} for more details). For this reason an $es$-expansion is preferable for numerical integrations as the Laplace coefficients depend only on the constant reference inclination and the semimajor axis ratio $\alpha$.  The version of the new disturbing function as an $e$-expansion is useful for analytical developments as we show in the next Section  with the derivation of mean motion resonance widths as a function of the asteroid's inclination.

\section{Resonance width at arbitrary inclination}\label{section5}
We illustrate the use of the disturbing function for arbitrary inclination by deriving the resonance width of a number of pure eccentricity mean motion resonances. Such resonances have arguments that do not involve the longitude of ascending node explicitly and consequently force amplitudes that depend exclusively on eccentricity. The classical disturbing function that has always been used for their study affirms that pure eccentricity resonances whose arguments for prograde motion may be written as $q\lambda-p\lambda^\prime+(p-q) \varpi$ do not depend on the orbital inclination. In effect,  only when the secular potential is added to the perturbing force,  does inclination affect the resonance albeit indirectly. Here we show using the new disturbing function that pure eccentricity resonances do intrinsically depend on the inclination for both prograde and retrograde motion. This helps us  demonstrate the importance of the new disturbing function in modelling three-dimensional resonance configurations accurately and caution against the general use of resonant dynamics properties obtained from the classical disturbing function that is based on the two-dimensional three-body problem. 

In order to derive resonance widths as functions of inclination, we upgrade the pendulum model of resonance and apply it to specific examples that include inner resonances with Jupiter and outer resonances with Neptune. To this end, in the next Section, we review the pendulum model then extend it to the presence of two harmonics so as to model the asymmetric librations of Neptune's resonances.

\subsection{Two-harmonics pendulum model of resonance}\label{ss5.1}
The pendulum model of resonance is the simplest method that allows the determination of the semimajor axis width of a mean motion resonance. Its drawback is the fact that it does not apply to orbits with very small eccentricity. The method is therefore adequate for our illustration needs as most asteroids, Centaurs and TNOs that interest us do not have negligible eccentricities. In the following, we briefly review how the pendulum model works and its relationship to the Poincar\'e model of resonance \citep{Poincare02}.  

Consider an asteroid in a $p$:$q$ resonance with the planet of  resonant argument $\phi_k^{p:q}=q\lambda-p\lambda^\prime-(q-p)\Omega-k\omega$.  We may write down the Lagrange planetary equation that governs the time evolution of the mean motion  $n=(GM_\star a^{-3})^{1/2}$ \citep{BC61} exactly as:
\begin{equation}
\dot n= -3 a^{-2} \partial_\sigma R= -\frac{3 n^2 \alpha \, m^\prime}{M_\star}\partial_\sigma \bar{R}.\label{dotn}
\end{equation}
 where $\sigma$ is related to the mean longitude at epoch $\tau=\lambda-n t$ by $\dot\sigma=t\dot n+ \dot \tau$  and $\bar{R}$ is the disturbing function with its direct, indirect and secular parts written as $\bar R=(f_{k,d}^{p:q} +f_{k,i}^{p:q}) \cos \phi_k^{p:q} + \bar{R}_s$ for the $p$:$q$ resonance. The force amplitudes  $f_{k,d}^{p:q}$ and $f_{k,i}^{p:q}$ include the eccentricity and inclination dependence of the resonant term. Next, consider the second derivative of the resonant argument with respect to time:
\begin{equation}
\ddot \phi_k^{p:q}= q(\dot n+\ddot\sigma)-(q-p)\ddot\Omega-k\ddot\omega.\label{ddotphi}
\end{equation}
The reference to the mean motion of the planet has disappeared as the asteroid is considered massless leaving the planet's orbit unperturbed. 

The working assumption of the pendulum model of resonance is to neglect{,} in the evolution of the resonant argument (\ref{ddotphi}), the second derivatives $\ddot\sigma$, $\ddot\Omega$, and $\ddot\omega$  {in comparison} to the time variation of the mean motion $n$. This  is generally valid unless the eccentricity is small {because} such angles as $\sigma$, $\Omega$, and $\omega$ would have rapid variations \citep{ssdbook}. {Following} the same logic, the secular term $\bar{R}_s$ has little effect on the resonance's dynamics as it does not influence the mean motion but concerns the discarded rates $\ddot\sigma$, $\ddot\Omega$, and $\ddot\omega$. The resonant state that the pendulum model describes is that of the stable branch of critical points  that extends to large eccentricity in the Poincar\'e model of resonance \citep{ssdbook}.  Since our main interest is the dynamics of  asteroids, Centaurs and TNOs whose orbits are not circular, we may use the pendulum model assumption to study the resonance's dynamics by substituting (\ref{dotn}) into (\ref{ddotphi}) to get:
\begin{equation}
\ddot \phi_k^{p:q}= \frac{3 n^2 q^2 m^\prime  \alpha}{M_\star}  (f_{k,d}^{p:q} +f_{k,i}^{p:q}) \ \sin \phi_k^{p:q}. \label{oneharmonic}
\end{equation}
This equation describes the evolution of the pendulum of angle $\phi_k^{p:q}$ and of natural frequency $|3 n^2 q^2 m^\prime M_\star^{-1}\alpha  (f_{k,d}^{p:q} +f_{k,i}^{p:q}) |^{1/2}$.  The sign of $f_{k,d}^{p:q} +f_{k,i}^{p:q}$ determines whether stable libration occurs around $\phi_k^{p:q}=0^\circ$ when it is negative or around  $\phi_k^{p:q}=180^\circ$ when it is positive. 

\begin{figure*}
\begin{center}
{ 
\includegraphics[width=52mm]{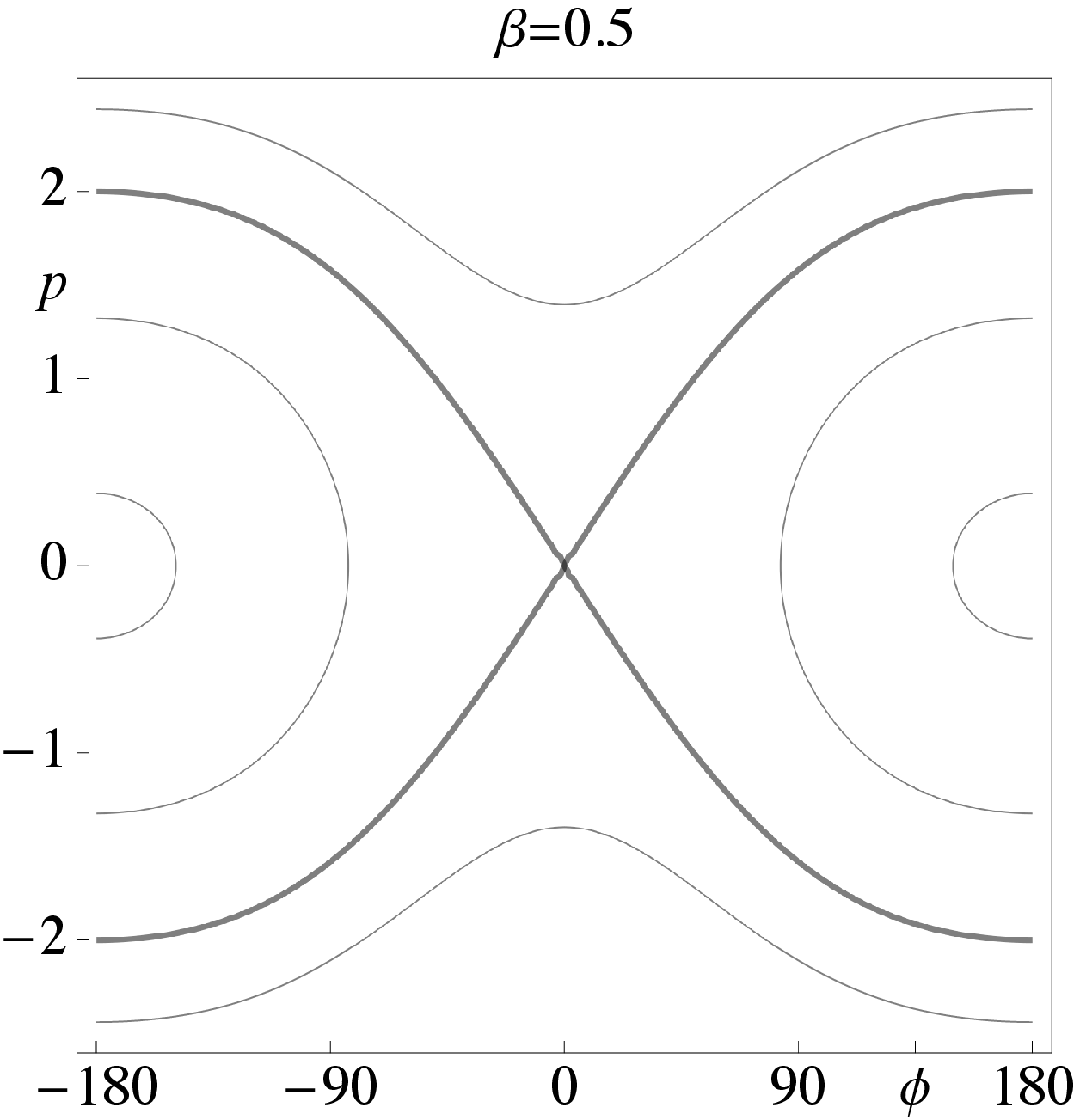}\hspace{10mm}\includegraphics[width=52mm]{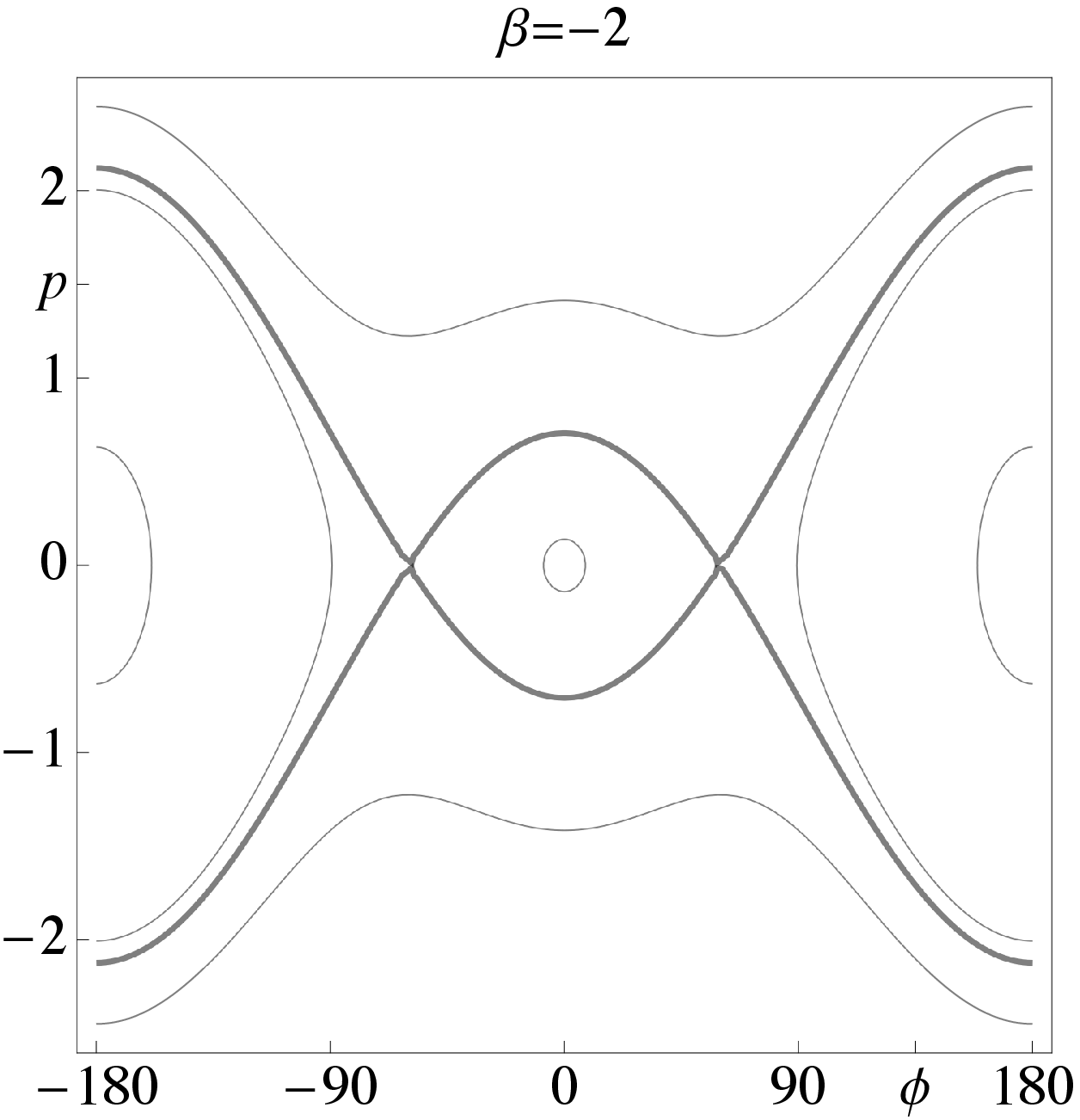}\hspace{10mm}\includegraphics[width=52mm]{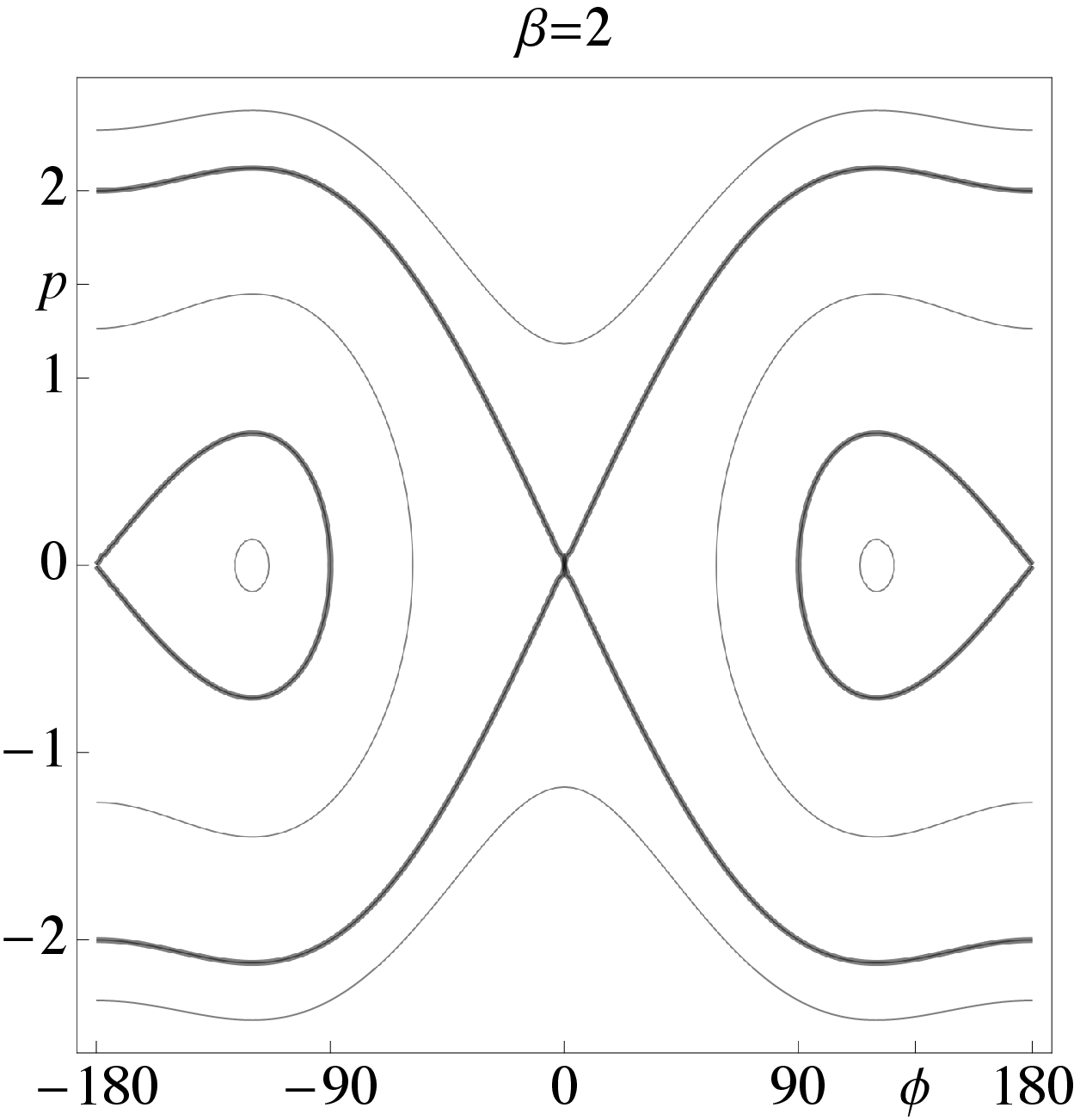}
}
\caption{Level curves of the two-harmonics pendulum Hamiltonian $H$ (\ref{hamil}). Separatrices are shown with bold lines.}\label{f1}
\end{center}

\end{figure*}

So far we have considered a single harmonic of the $p$:$q$ resonance given by the angle $\phi_k^{p:q}$. Modelling resonances with a single harmonic whether through the pendulum model or the Poincar\'e Hamiltonian is known to be  sufficient for inner resonances. {However, for outer resonances,} higher harmonics' amplitudes modify the topology of phase space giving rise to asymmetric librations that depend on the orbit's eccentricity \citep{Malhotra96,WinterMurray97} (we will show in Section \ref{ss5.3} that they are  intrinsically dependent on inclination as well).  This behaviour affects particularly outer resonances of the type 1:$q$ \citep{Bruno}. For such resonances, we upgrade the pendulum model by including the second harmonic $\phi_{2k}^{2p:2q}=2\phi_k^{p:q}$ in a way similar to Andoyer's Hamiltonian model \citep{Andoyer02,Beauge94}. The same restriction on the one-harmonic  model applies here in that very small eccentricities may not be studied by the two-harmonics pendulum model. For outer resonances, the two-harmonics pendulum equation is given as:
\begin{eqnarray}
\ddot \phi_k^{p:q}&=& \frac{3 n^2 q^2 m^\prime  \alpha}{M_\star} \left[ (f_{k,d}^{p:q} +f_{k,i}^{p:q}) \ \sin \phi_k^{p:q}\right. \nonumber 
\\&&\ \ \ \ \ \ \ \ \  \left. +2 (f_{2k,d}^{2p:2q} +f_{2k,i}^{2p:2q}) \ \sin 2 \phi_k^{p:q}\right]\label{twoharmonic}.
\end{eqnarray}
In order to derive the properties of this dynamical system, we write it in the following way $\ddot \phi = u \sin \phi + v \sin 2\phi$.
We may choose  $u>0$ as when it is negative, the variable change $\phi\rightarrow \phi+180^\circ$ makes the coefficient of $\sin\phi$ positive. With this convention, stable librations of the first harmonic ($v=0$) occur around $180^\circ$. Next, we change the time variable to $t^\prime = |u|^{1/2} t$.  The reduced two-harmonics pendulum equation then becomes $\phi^{\prime\prime}= \sin \phi (1+\beta \cos\phi)$ where  $\phi^{\prime\prime}$ denotes the second  derivative with respect to $t^\prime$ {and:}
\begin{equation}\beta= \frac{2v}{|u|}=\frac{4(f_{2k,d}^{2p:2q} +f_{2k,i}^{2p:2q})}{|f_{k,d}^{p:q} +f_{k,i}^{p:q}|}.\label{beta}\end{equation} 
The corresponding Hamiltonian is given as:
 \begin{equation}H=\frac{p^2}{2} +\cos\phi +\frac{\beta}{4}\, \cos(2\phi),\label{hamil}\end{equation}
 where the momentum $p=\phi^\prime$.  The critical points of this system are given by $p=0$ and $\sin\phi=0$ and when possible by the {asymmetric centres:}
 \begin{equation}
 \phi_a=\pm\arccos(-\beta^{-1}). \label{alib}
 \end{equation}
 The Hamiltonian's level curves are shown in Figure (\ref{f1}) for three values of the parameter $\beta$. 

If $|\beta|<1$, the position and stability of the two equilibrium points of the one-harmonic pendulum model  are unchanged by the addition of the second harmonic. The motion separatrix passes through the unstable point $\phi=0^\circ$ (as $u>0$) and has an energy $H=1+\beta/4$. Therefore the corresponding resonance halfwidth is given as $p(\phi=180^\circ)=2$ which in fully dimensional units reads $\dot\phi=2|u|^{1/2}$ the same as that of the one-harmonic pendulum model ($\beta=0$). Hence if the second harmonic is weaker than a quarter of the first, libration stability and resonance width are unchanged with respect to the one-harmonic pendulum model. The modified properties are the libration time and elongation of the angle $\phi$. 

When the second harmonic is stronger than a quarter of the first, $|\beta|\geq1$,  two possibilities must be considered $\beta<0$ and $\beta>0$. For the former ($\beta\leq-1$), librations around $\phi=180^\circ$ remain stable as can be seen directly from the reduced two-harmonics pendulum equation but in addition to them, stable librations around $\phi=0^\circ$ are now possible. Two unstable points form at $\phi_{a,u\pm}$ (\ref{alib}). As the separatrix now passes through these points, the corresponding energy is given as $H=-\beta/4-1/2\beta$. This enables us to determine the maximum half-widths of stable librations around $0^\circ$ and $180^\circ$ as  $p(\phi=0^\circ)= |\beta|^{-1/2}|\beta+1|$  and $p(\phi=180^\circ)=|\beta|^{-1/2}|\beta-1| $. 
 
When $\beta\geq 1$, both one-harmonic critical points at $\phi=0^\circ$ and $180^\circ$ become unstable and stable asymmetric librations appear around $\phi_{a,s\pm}$ (\ref{alib}). There are two separatrices in this case. One that goes through $\phi=180^\circ$ and defines the maximum limit for asymmetric librations for an energy $H=\beta/4-1$. The second separatrix passes through $\phi=0^\circ$ and delineates the domain of symmetric libration around both points $\phi_{a,s\pm}$ (\ref{alib}) from the  general circulation domain; its energy is  $H=\beta/4+1$. The asymmetric libration halfwidth is $p(\phi_{a,s\pm})= |\beta|^{-1/2}|\beta-1|$  whereas that around both equilibria is $p(\phi_{a,s+} {\rm \ and\ } \phi_{a,s-})=|\beta|^{-1/2}|\beta+1| $.

In order to find the resonance widths in terms of the semimajor axis, we recall that in the pendulum approximation $\dot n = q^{-1}\ddot \phi$. This leads to $n^\prime = q^{-1}|u|^{1/2} \phi^{\prime\prime}$.  Eliminating the time variable $t^\prime$ in favour of $\phi$ leads to $dn/d\phi = q^{-1}|u|^{1/2} \phi^{\prime\prime}/ \phi^\prime$. Using the energy integral $H$ and integrating\footnote{Denoting $V(\phi)$ the perturbing potential $\cos\phi +{\beta}\, \cos(2\phi)/4$, we have $\phi^{\prime}=[2(H-V)]^{1/2}$ and $\phi^{\prime\prime}=-dV/d\phi$. The quantity $dn/d\phi=-|u|^{1/2}q^{-1}(dV/d\phi)/[2(H-V)]^{1/2}$.} with respect to $\phi$  yields the general expression of the mean motion as a function of $\phi$ as:
\begin{equation}n=q^{-1}|u|^{1/2} [2 (H-\cos \phi -\beta\cos (2\phi)/4)]^{1/2}.\end{equation} 
The general expression of the mean motion can be used to derive the resonance width by applying it to a given separatrix defined by its energy and the unstable points it passes through and combining it with the Kepler's third law $\Delta a =2(3n)^{-1}\Delta n\, a$. Equivalently, noting that $dn=q^{-1}|u|^{1/2} dp$, the mean motion widths are proportional to the momentum widths derived previously through $\Delta n=q^{-1}|u|^{1/2} \Delta p$. Applying Kepler's third law yields the resonance widths in terms of semimajor axis.

Thus when the second harmonic is weaker than a quarter of the first, $|\beta|<1$, stable libration occurs around $\phi=180^\circ$ with $H=1+\beta/4$ and the semimajor axis resonance width is given as:
\begin{equation}
\Delta_0 a_k^{p:q} = \frac{4|u|^{1/2}a_k^{p:q}}{3qn} =\left[\frac{16\, \alpha\, m^\prime|f_{k,d}^{p:q} +f_{k,i}^{p:q}|}{3M_\star}\right]^\frac{1}{2} a_k^{p:q}\label{reswidth0}.
\end{equation}
where $a_k^{p:q}=\alpha a^\prime$ is the resonant semi-major axis.   
This is the classical formula of the one-harmonic pendulum model \citep{ssdbook} that is used for the critical arguments obtained from the classical disturbing function ($I_r=0^\circ$) to visualise the resonance width as a function of eccentricity, assess how it increases with eccentricity and determine the possible resonance overlapping that triggers chaotic motion.

When the second harmonic is stronger than a quarter of the first and its amplitude is negative, $\beta\leq-1$, stable librations may occur around $\phi=0^\circ$ and $180^\circ$, with a larger width for the latter.  There is a single separatrix corresponding to the energy $H=-\beta/4-1/2\beta$ and  the semimajor axis width at $\phi=0^\circ$ is given as:
\begin{eqnarray}
\Delta_1 a_k^{p:q} \!\!\!\!\!\!&=& \!\!\!\!\frac{2|2v +|u||a_k^{p:q}}{3qn|2v|^{1/2}}, \label{reswidth1}\\
\Delta_1 a_k^{p:q} \!\!\!\!\!\!&=& \!\!\!\!\!\!\!\left[\frac{ \alpha\, m^\prime}{3M_\star}\right]^\frac{1}{2}  \!\!\frac{|4(f_{2k,d}^{2p:2q} +f_{2k,i}^{2p:2q})+|f_{k,d}^{p:q} +f_{k,i}^{p:q}||}{|f_{2k,d}^{2p:2q} +f_{2k,i}^{2p:2q}|^{1/2}}a_k^{p:q}.\nonumber
\end{eqnarray}
Libration around $\phi=180^\circ$ occurs with maximum semimajor axis width:
\begin{eqnarray}
\Delta_2 a_k^{p:q} \!\!\!\!\!\!&=& \!\!\!\!\frac{2|2v -|u||a_k^{p:q}}{3qn|2v|^{1/2}}, \label{reswidth2}\\
\Delta_2 a_k^{p:q} \!\!\!\!\!\!&=& \!\!\!\!\!\!\!\left[\frac{ \alpha\, m^\prime}{3M_\star}\right]^\frac{1}{2}  \!\!\frac{|4(f_{2k,d}^{2p:2q} +f_{2k,i}^{2p:2q})-|f_{k,d}^{p:q} +f_{k,i}^{p:q}||}{|f_{2k,d}^{2p:2q} +f_{2k,i}^{2p:2q}|^{1/2}}a_k^{p:q}.\nonumber
\end{eqnarray}

When the second harmonic is stronger than a quarter of the first and its amplitude is positive, $\beta\geq1$, asymmetric librations at either $\phi_{s\pm}=\pm\arccos(-\beta^{-1})$ are possible and are delineated by a separatrix of energy $H=\beta/4-1$, the resonance's semimajor axis width is given by $\Delta_2 a_k^{p:q}$. A second separatrix is associated with symmetric librations about both $\phi_{s\pm}$ centred around $\phi=180^\circ$ and of energy $H=\beta/4+1$. The corresponding resonance width is given by $\Delta_1 a_k^{p:q}$. 

All this analysis was done for a positive first harmonic amplitude ($f_{k,d}^{p:q} +f_{k,i}^{p:q}>0$). If that amplitude is negative then the only change that occurs is that all critical points have to be shifted by $180^\circ$ as the variable change $\phi=180^\circ+\bar\phi$ produces a two-harmonics pendulum equation for  $\bar \phi$ identical to that of $\phi$ except for a positive sign for the  first harmonic coefficient. Thus, for instance, stable asymmetric librations for $f_{k,d}^{p:q} +f_{k,i}^{p:q}<0$ {occur at} $\phi_{a,s\pm}=180^\circ\pm\arccos(-\beta^{-1})$ when $\beta\geq 1$.  Examples of the use of the two-harmonics pendulum  are given is Section \ref{p5.3.2} for Neptune's outer 1:2 and 1:3 resonances.

\subsection{Prograde and retrograde resonant arguments}\label{ss5.2}
The disturbing function of an asteroid that interacts with a planet on a circular orbit involves the argument $\phi_k^{p:q}=q\lambda-p\lambda^\prime-(q-p)\Omega-k\omega$ with the usual orbital angles defined for prograde motion. To emphasise 
the prograde nature of the chosen angles, we write the argument as a function of the longitude of pericentre, $\varpi$, instead of the argument of pericentre, $\omega$, as the former's definition changes when motion is retrograde:
\begin{equation}
\phi_{+}= q\lambda-p\lambda^\prime-k \varpi +(p-q+k)\Omega. \label{prophi}
\end{equation}
For a given ($p$,$q$,$k$) integer set, we can follow the dependence of the corresponding resonance width on the reference inclination even when  $I_r$ exceeds $90^\circ$  as it is known that libration in prograde resonance exists at retrograde inclinations  albeit with a modified dynamical behaviour. Such libration was observed  when retrograde motion is about to be captured in mean motion resonance  in a universal process termed the three-stage capture process of high order resonance \citep{NamouniMorais15}. It may be summed up briefly as follows: capture first occurs in the prograde resonant argument that behaves like an inclination resonance. Motion enters the second stage when the Kozai-Lidov resonance is activated followed by the final capture in the retrograde  resonance termed the third stage. Examples of this mechanism are given in  \citep{NamouniMorais15} for 1:2  resonance and an inclination $I=170^\circ$ and in \citep{MoraisNamouni16} for the  1:1 resonance with an inclination $I=179.999^\circ$. 

We may also use the argument $\phi_k^{p:q}$ to study retrograde resonance  by adopting the angles $\lambda^\star$ and $\varpi^\star$ that are compatible with retrograde motion and related to   the mean longitude through $\lambda=\lambda^\star+2\Omega$ and to the longitude of perihelion through $\varpi=\varpi^\star+2\Omega$ \citep{MoraisNamouni13a}. The retrograde resonant argument thus reads:
\begin{equation}
\phi_{-}= q\lambda^\star-p\lambda^\prime-k \varpi^\star +(q+p-k)\Omega. \label{retphi}
\end{equation}
It is evident from the expressions (\ref{prophi}) and (\ref{retphi}) that what defines the prograde or retrograde flavour of a $p$:$q$ resonance is the choice of the integer $k$. For instance, if we are interested in the prograde pure eccentricity resonant term (\ref{prophi}) then $k=q-p$ whereas if we go for the equivalent retrograde pure eccentricity resonant term (\ref{retphi}), $k=q+p$. In this regard, we remark that {for} prograde pure eccentricity arguments, {the} integer $k$ depends on  the perturber's location: inside perturbers have positive $k$ whereas outside perturbers  have negative $k$. Retrograde pure eccentricity arguments do not distinguish between outside and inside perturbers as $k$ is always positive. In the following, we estimate prograde as well as retrograde resonance widths.

\subsection{Resonance widths}\label{ss5.3}
The resonance width  is a function of  eccentricity $e$, inclination $s$ and reference inclination $I_r$ through the force amplitudes $f_{k,d}^{p:q}$ and $f_{k,i}^{p:q}$ (\ref{reswidth0},\ref{reswidth1},\ref{reswidth2}).  In this illustration of the use of the new disturbing function, we seek to understand how pure eccentricity resonance widths depend on the inclination of the resonant motion of an asteroid. The classical disturbing function being a series expansion with respect to circular co-planar orbits, predicts that pure eccentricity terms do not depend on the asteroid's orbital inclination. They may depend indirectly on the inclination variations only through the secular part of the disturbing function that couples the evolution of eccentricity to that of inclination but the inner workings of such resonances remain inclination-free. The fact that the force amplitudes of the new disturbing function depend on the reference inclination $I_r$ through the two-dimensional Laplace coefficients will invalidate this conclusion. 

As  we  need to select a meaningful reference inclination $I_r$ for a resonant asteroid with an inclination $I$, we can invoke the two interpretations of the new disturbing functions to this effect. If we choose to cancel the free inclination variable through $s\equiv 0$ and retain only the eccentricity expansion, then the reference inclination in the disturbing function is just  the asteroid's inclination and is present in the two-dimensional Laplace coefficients.  Since the resonance width formulae in the pendulum approximation assume exact resonance for $e$ and $I$, these elements are actually constant. With the interpretation of new disturbing function as a double $es$--expansion, 
 the reference inclination can be set to the asteroid's mean inclination $I_r=\langle I\rangle$, {that is its value after short period variations are averaged out. More precisely, the reference inclination can be set to the asteroid's proper inclination \citep{Knezevic00}.} Choosing  those orbits whose inclination amplitude $s$ is small compared to the mean or {proper} value  $\langle I\rangle$ gives us the resonance widths of the previous choice with $s\equiv 0$. Therefore in the context of the pendulum model of resonance, whether the reference inclination is interpreted as the asteroid's inclination, {or the proper value,} the derived resonance widths  are identical.
 
For the asteroid's orbital eccentricity, we choose three values $e=0.1$, $e=0.3$ and $0.5$. The classical studies of resonance dynamics indicate that  resonance width is an increasing function of eccentricity when orbits are not nearly circular \citep{ssdbook}.

\subsubsection{Inner 2:1 and 3:1 Jupiter eccentricity resonances}\label{p5.3.1}
The appropriate resonant argument for the pure eccentricity prograde inner 2:1 resonance is obtained with $k=-1$ as $\phi_{-1}^{2:1}=\lambda - 2\lambda^\prime +\varpi$. As the inclination's oscillations are assumed negligible ($s\ll 1$), we neglect all inclination terms in the force amplitudes (proportional to powers of $s$). To fourth order in eccentricity, the amplitude  thus involves the coefficients $c^1_{10}(2,-1,\alpha=2^{-2/3},I_r+180^\circ)$ and  $c^1_{30}(2,-1,\alpha=2^{-2/3},I_r+180^\circ)$ in Table \ref{t2}  where we used the relationship (\ref{cmkpk}) given in Section \ref{section3} between coefficients $c^k_{mn}$ of positive and negative $k$. The force amplitude expression reads:
\begin{eqnarray}
f_{-1,d}^{2:1}&=& -\frac{e}{4} (4 \bar A_{0, 2, -2, 0} + \bar A_{0, 2, -2, 1})\nonumber \\&& +\frac{e^3}{32} (28 \bar A_{0, 2, -2, 0} + 
   5 \bar A_{0, 2, -2, 1}-6 \bar A_{0, 2, -2, 2} \label{force2t1}\nonumber\\&&
    - \bar A_{0, 2, -2, 3}),
\end{eqnarray} 
where $\bar A_{i,j,k,l}(I_r)= A_{i,j,k,l}(I_r+180^\circ)$. We may remove the negative signs and the bar from the previous expression as  $\bar A_{i,j,-k,l}(I_r)= \bar A_{i,j,k,l}(I_r+180^\circ)=A_{i,j,k,l}(I_r+360^\circ)=A_{i,j,k,l}(I_r)$ (see equation \ref{ALap2}).  For inner resonances, there is no contribution from the  indirect part of the disturbing function. 

Similarly, the resonant argument for the pure eccentricity prograde inner 3:1 resonance is obtained with $k=-2$ as $\phi_{-2}^{3:1}=\lambda - 3\lambda^\prime +2\varpi$. The corresponding force amplitude to fourth order  in eccentricity are obtained from the relationship (\ref{cmkpk}) as $c^2_{20}(3,-1,\alpha=3^{-2/3},I_r+180^\circ)$ and  $c^2_{40}(3,-1,\alpha=3^{-2/3},I_r+180^\circ)$ (Table \ref{t3}): 
\begin{eqnarray}
f_{-2,d}^{3:1}&=& 
\frac{e^2}{16}(21 A_{0, 3, 3, 0} + 10 A_{0, 3, 3, 1} + A_{0, 3, 3, 2}) \label{force3t1}\nonumber\\&&
+\frac{e^4}{192} (15 A_{0, 3, 3, 2} -186 A_{0, 3, 3, 0} - 
   122 A_{0, 3, 3, 1} \nonumber\\&&+
   12 A_{0, 3, 3, 3}+ A_{0, 3, 3, 4}),
\end{eqnarray} 
where like the 2:1 expression we removed the bar and negative signs in $A_{i,j,k,l}$ and set their inclinations to $I_r$ instead of $I_r+180^\circ$. 

In order to estimate the resonance's width, we use the one-harmonic pendulum model as it is most appropriate for inner resonances. Throughout this Section and the next, the values of the inclination-dependent two-dimensional  Laplace coefficients are determined using the {series} derived in Appendix \ref{app} (\ref{bsjk}) with an $\alpha$-expansion order $N_\alpha=20$. Figure (\ref{f2}) shows the resonance widths obtained by inserting the expressions (\ref{force2t1}) and  (\ref{force3t1}) into the resonance width formula (\ref{reswidth0}) as a function of the asteroid's inclination and the three possible eccentricities 0.1, 0.3 and 0.5.  Jupiter's parameters are semimajor axis $a^\prime=5.2$\, AU, mass $m^\prime/M_\star=10^{-3}$. The widths of prograde inner resonances are found to be monotonically decreasing  functions of inclination that vanish for exactly retrograde coplanar motion. The eccentricity dependence agrees with the classical result that  larger eccentricity implies larger width.

The appropriate resonant arguments for the pure eccentricity retrograde inner 2:1 and 3:1 resonances are obtained with $k=3$ and $k=4$ respectively as  $\phi_{3}^{2:1}=\lambda^\star - 2\lambda^\prime -3\varpi^\star$ and $\phi_{4}^{3:1}=\lambda^\star - 3\lambda^\prime -4\varpi^\star$. The corresponding force amplitudes to fourth order in eccentricity are found in Tables \ref{t4} and \ref{t5} for the direct part  $c^3_{30}(2,1,\alpha=2^{-2/3},I_r)$ and $c^4_{40}(3,1,\alpha=3^{-2/3},I_r)$ as:
\begin{eqnarray}
f_{3,d}^{2:1}&=&
\frac{e^3}{96} (4 A_{0, 2, -2, 0}- 3 A_{0, 2, -2, 1} -6 A_{0, 2,-2, 2} \nonumber \\&& - A_{0, 2, -2, 3}), \label{force2t1r}\\
f_{4,d}^{3:1}&=& 
\frac{e^4}{768} ( 30 A_{0, 3, -3, 2}-15 A_{0, 3, -3, 0}  -4A_{0, 3, -3, 1}\nonumber\\&& + 12 A_{0, 3, -3, 3}  + A_{0, 3, -3, 4}),\label{force3t1r} 
\end{eqnarray} 
where unlike the prograde expressions (\ref{force2t1},\ref{force3t1}) we may not remove the negative signs in $A_{i,j,k,l}$ unless we change  the inclination from $I_r$ to $I_r+180^\circ$ (see equation \ref{ALap2}). It is interesting to note that the expressions (\ref{force2t1},\ref{force3t1},\ref{force2t1r},\ref{force3t1r}) are identical to those obtained with nearly coplanar motion, that is $j=k$ in $A_{i,j,k,l}$  for prograde motion and $j=-k$ for retrograde motion, except for the fact that the new Laplace coefficients in $A_{i,j,k,l}$   now depend on the reference  inclination. 
The widths of the 2:1 and 3:1 retrograde Jupiter resonances obtained from expression (\ref{reswidth0}) are shown in Figure (\ref{f2}) where they are found to be naturally narrower than their prograde counterparts. The retrograde 2:1 resonance width does not decrease monotonically and has a maximum near $\sim 40^\circ$ whereas both retrograde resonance widths vanish for exact prograde coplanar motion. 

We therefore conclude that pure eccentricity inner resonances whether prograde or retrograde depend intrinsically on the asteroid's inclination and therefore will influence the resonance's capture efficiency.  This conclusion is contrary to what may be inferred from the classical disturbing function whose reference inclination is $I_r=0^\circ$.

\subsubsection{Outer 1:2 and 1:3 Neptune eccentricity resonances}\label{p5.3.2}
In the two-dimensional three-body problem, outer resonances of the type 1:$q$ are known to exhibit asymmetric librations whose resonant arguments' equilibria lie somewhere between $50^\circ$ and $180^\circ$  depending on the asteroid's orbital eccentricity \citep{Bruno,Beauge94,Malhotra96,WinterMurray97}. The origin of such equilibria for a $p$:$q$ resonance of  argument $\phi_k^{p:q}$ is related to the importance of the second harmonic's  $\phi_{2k}^{2p:2q}$ amplitude.\footnote{Historically, it was thought that at least nine harmonics were required to establish asymmetric librations \citep{Message58}.}   For this reason, we developed the new two-harmonics pendulum model  and proceed to apply it to Neptune's outer resonances. 

The pure eccentricity prograde outer 1:2 resonant argument is obtained with $k=1$ as $\phi_1^{1:2}=2 \lambda - \lambda^\prime -\varpi$. The corresponding force amplitudes to fourth order in eccentricity are found in Table \ref{t2} for the direct part: $c^1_{10}(1,2,\alpha=2^{2/3},I_r)$ and $c^1_{30}(1,2,\alpha=2^{2/3},I_r)$, and in Table \ref{t6} for the indirect part. The force amplitude of the second harmonic  $\phi_2^{2:4}=4 \lambda - 2\lambda^\prime -2\varpi$ is read off Table \ref{t3}   ($k=2$)  with the coefficients  $c^2_{20}(2,4,\alpha=2^{2/3},I_r)$ and  $c^2_{40}(2,4,\alpha=2^{2/3},I_r)$ (no indirect terms are applicable). We thus have: 
\begin{eqnarray}
f_{1,d}^{1:2}&=& \frac{e}{4} (2A_{0, 1, 1, 0} - A_{0, 1, 1, 1})+ \nonumber \\ &&\frac{e^3}{32} ( 
   14 A_{0, 1, 1, 1}
 -20 A_{0, 1, 1, 0}  - A_{0, 1, 1, 3}),\\
f_{1,i}^{1:2}&=&  \frac{e \alpha}{16}   ( 3 e^2 -4) (1+\cos I_r),\\
f_{2,d}^{2:4}&=&
\frac{e^2}{16} (26 A_{0, 2, 2, 0}- 10 A_{0, 2, 2, 1}+ 
   A_{0, 2, 2, 2})+\nonumber\\&&
\frac{e^4}{192} ( 428 A_{0, 2, 2, 1}-1036 A_{0, 2, 2, 0}- 
   30 A_{0, 2, 2, 2}\nonumber\\&&- 8 A_{0, 2, 2, 3}+ 
   A_{0, 2, 2, 4}).
\end{eqnarray} 
Although there is no explicit inclination  dependence in the direct force amplitude, we remind the reader that the coefficients $A_{i,j,k,l}$ (\ref{Aijkl}) are functions of inclination through the two-dimensional Laplace coefficients $b^{jk}_s$ (\ref{Lap0}) like in the previous examples of Jupiter's inner resonances. The amplitudes are evaluated numerically for Neptune with a semimajor axis $a^\prime=30.11$\, AU, and mass $m^\prime/M_\star=5.12\times 10^{-5}$.

When estimating the resonance width with the two-harmonics model, two quantities must be monitored the ratio $\beta=4f_{2,d}^{2:4}/|f_{1,d}^{1:2}+f_{1,i}^{1:2}|$ (\ref{beta}) of the second harmonic's amplitude to a quarter of the first harmonic's absolute amplitude, as well as the sign of the first harmonic's amplitude  ${\rm sgn}_{1}^{1:2}={\rm sign}(f_{1,d}^{1:2}+f_{1,i}^{1:2})$. 

\begin{figure*}
\begin{center}
{ 
\includegraphics[width=75mm]{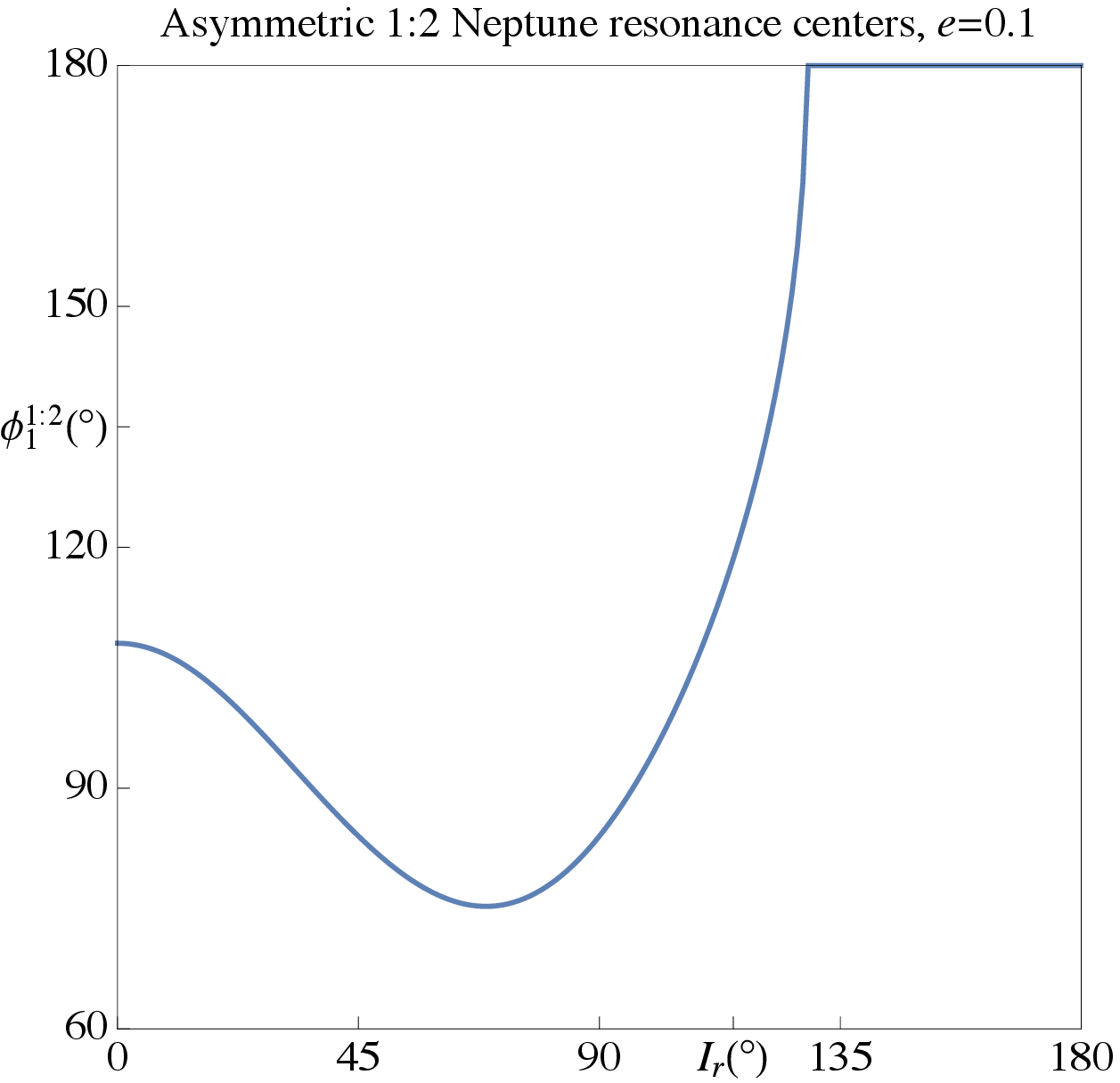}\hspace{10mm}\includegraphics[width=75mm]{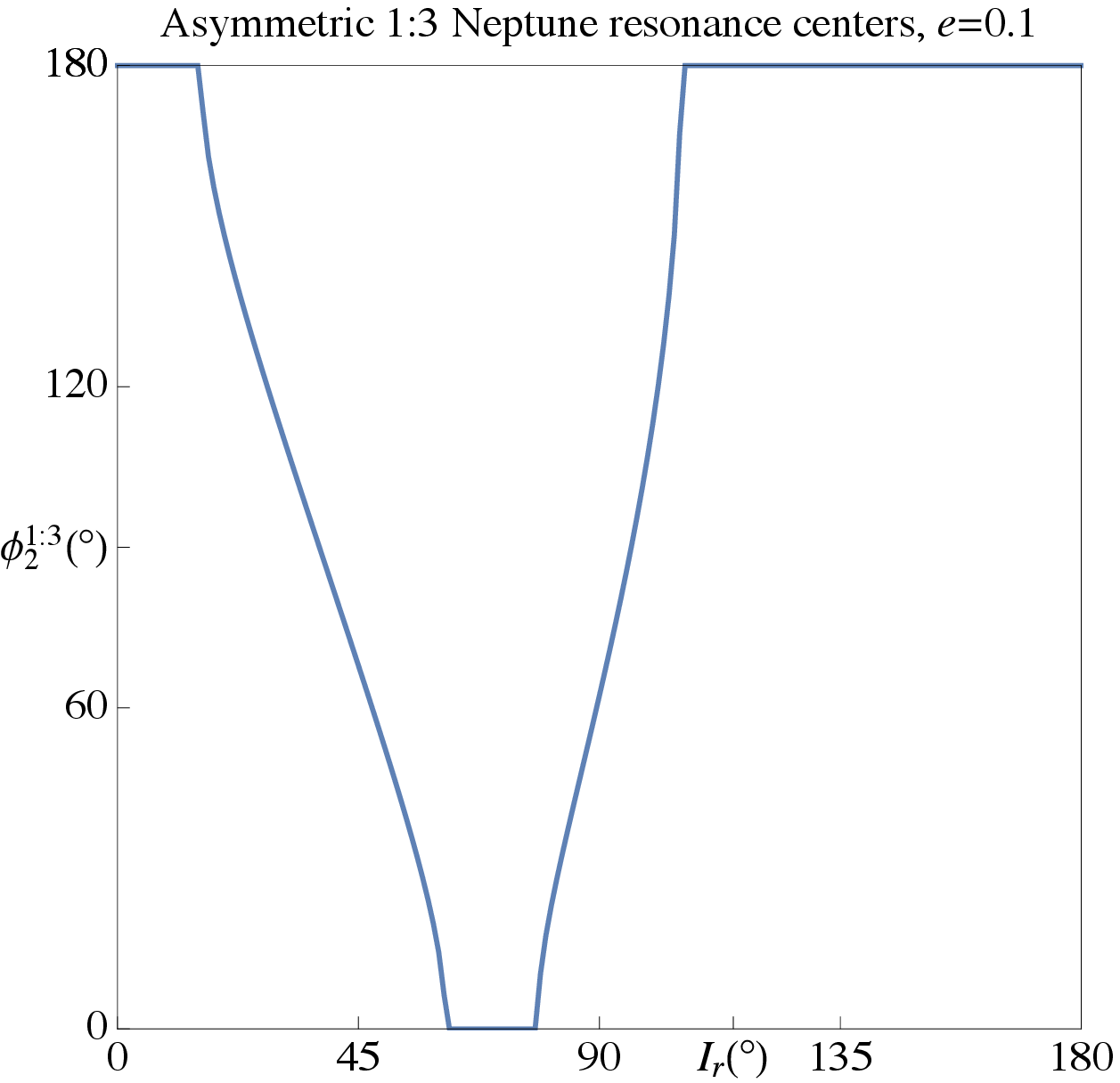}\\[5mm]
\includegraphics[width=75mm]{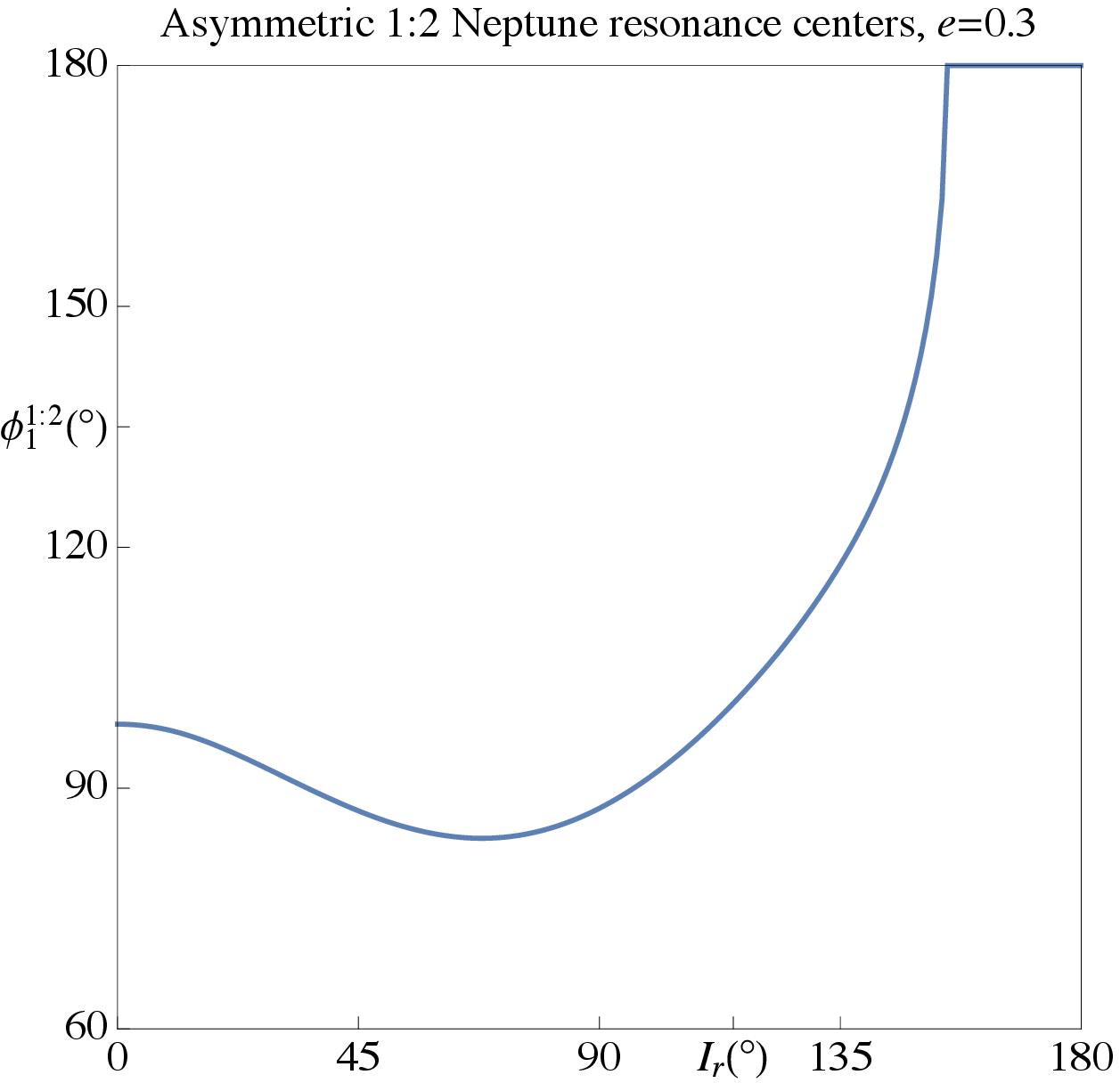}\hspace{10mm}\includegraphics[width=75mm]{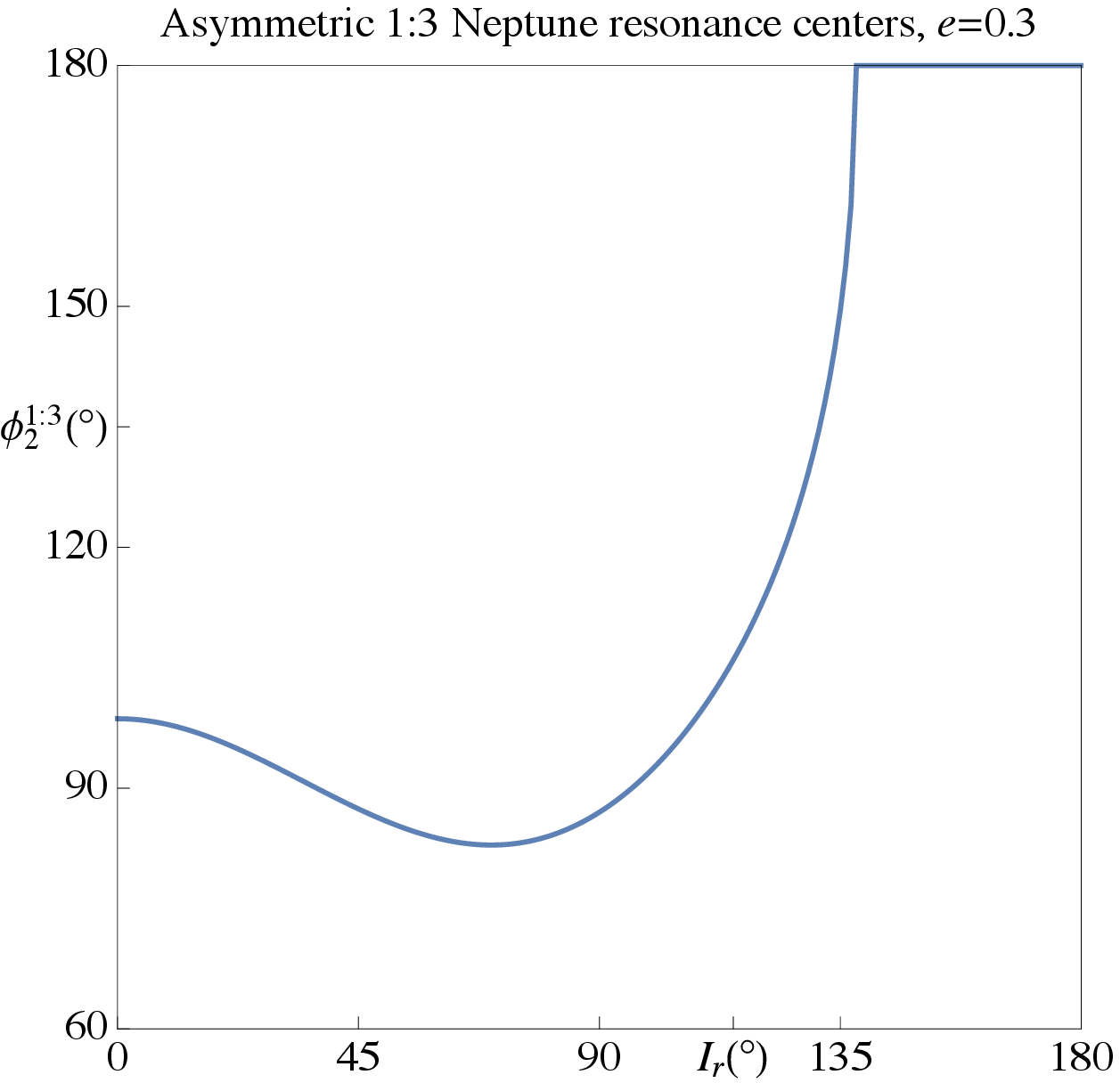}}
\caption{Outer 1:2 and 1:3 Neptune positive asymmetric resonance centres  (\ref{alib}) as function of the asteroid's average inclination $I_r$ for two eccentricity values $e=0.1$ and 0.3. }\label{f4}
\end{center}
\end{figure*}

{The} 1:2 resonance's $\beta$ is found to be always positive.  For $e=0.1$, the ratio $\beta \geq 1$ for $0^\circ\leq I_r \leq 129^\circ$ implying the presence of asymmetric librations up to that inclination and simple $180^\circ$-libration beyond it. The sign ${\rm sgn}_{1}^{1:2}< 0$ in the inclination interval $[37^\circ,96^\circ]$ and positive otherwise. The sign change allows the asymmetric librations to exist below the value $\phi_1^{1:2}=90^\circ$ that is reached at the edges of that interval where the separatrices of asymmetric librations and $180^\circ$ librations intersect. The resonance widths are thus obtained from equations (\ref{reswidth1} and \ref{reswidth2}) for $I_r\leq 129^\circ$ and from equation (\ref{reswidth0}) beyond that value. The resulting resonant widths as function of the inclination are shown in Figure (\ref{f3}). 
 It is found that the resonance width of asymmetric libration shows a slight increase starting from the planar configuration to reach a maximum at $36^\circ$ whereas the $180^\circ$-libration about both asymmetric centres increases and reaches a minimum at that inclination. A second extremum is reached at $96^\circ$ after which asymmetric libration width decreases until it disappears in favour of simple $180^\circ$-libration ($0<\beta<1$).
Figure (\ref{f4}) shows the location of the positive asymmetric libration centre (\ref{alib}); the {negative} one {is} symmetric with resect to $\phi_1^{1:2}=0^\circ$.  The asymmetric libration centre starts out at $\phi_1^{1:2}=108^\circ$ for $I_r=0^\circ$, reaches a minimum $\phi_1^{1:2}=75^\circ$ for $I_r=69^\circ$ to increase steadily to $180^\circ$ and disappears at $I_r= 129^\circ$. 

Increasing the eccentricity value to $e=0.3$ increases the resonance widths and produces a similar behaviour except  the asymmetric libration width that decreases as the asteroid's orbit leaves the planet's orbital plane. The three critical inclinations are $I_r=36^\circ$, $96^\circ$ where the libration centres reach $90^\circ$ and  $155^\circ$ beyond which asymmetric librations no longer exist. The location of asymmetric librations at $I_r=0^\circ$ is $\phi_1^{1:2}=98^\circ$
 and the  minimum of $83^\circ$ is reached at  $I_r=68^\circ$. Regardless of eccentricity, all resonant librations disappear for exactly coplanar retrograde motion $I_r=180^\circ$. 

We now examine the pure eccentricity prograde outer 1:3 resonance. Its argument is obtained with  $k=2$ as $\phi_2^{1:3}=3 \lambda - \lambda^\prime -2\varpi$ and the corresponding force amplitudes are found in Table \ref{t3} for the direct part: $c^2_{20}(1,3,\alpha=3^{2/3},I_r)$, $c^2_{40}(1,3,\alpha=3^{2/3},I_r)$, and in Table \ref{t6} for the indirect part. The second harmonic  $\phi_4^{2:6}=6 \lambda - 2\lambda^\prime -4\varpi$ has the $k=4$-amplitude coefficient $c^4_{40}(2,6,\alpha=3^{2/3},I_r)$ found in Table \ref{t5} (no indirect terms are applicable). The force amplitudes are given explicitly as:
\begin{eqnarray}
f_{2,d}^{1:3}&=& 
\frac{e^2}{16}(9 A_{0, 1, 1, 0} -6 A_{0, 1, 1, 1} + A_{0, 1, 1, 2})\nonumber\\&&+
\frac{e^4}{192} (126 A_{0, 1, 1, 1}-162 A_{0, 1, 1, 0} -21 A_{0, 1, 1, 2} \nonumber\\&& -4 A_{0, 1, 1, 3}+ A_{0, 1, 1, 4})
,\\
f_{2,i}^{1:3}&=&\frac{3e^2\alpha}{16}   ( e^2 -1)  (1+ \cos I_r ),\\
f_{4,d}^{2:6}&=&\frac{e^4}{768} (2760 A_{0, 2, 2, 0}- 1464 A_{0, 2, 2, 1} 
   +300 A_{0, 2, 2, 2}\nonumber\\&&- 28 A_{0, 2, 2, 3}+ 
   A_{0, 2, 2, 4}).
\end{eqnarray} 

For the planar 1:3 resonance, asymmetric librations are triggered at  a larger eccentricity value than the 1:2 resonance, a  threshold estimated numerically at $e\sim 0.13$ \citep{Beauge94,Malhotra96}. It is therefore not surprising that for an eccentricity $e=0.1$, the ratio $0<\beta<1$ and asymmetric librations are absent for $I_r=0^\circ$ (Figure \ref{f3}). At the same eccentricity but for three-dimensional motion, two small asymmetric libration islands appear in the inclination intervals $[15^\circ,62^\circ]$ and $[78^\circ,106^\circ]$ (Figure \ref{f3}). The position of the positive libration centre decreases with inclination from  $180^\circ$ to $0^\circ$  in the first interval and increases with inclination from  $0^\circ$ to $180^\circ$. In the same inclination intervals, $180^\circ$-libration round both asymmetric centres is naturally possible. Otherwise all libration occurs around the centre defined by the first harmonic namely $180^\circ$ (as $0<\beta\leq1$). For the larger eccentricity $e=0.3$, asymmetric librations exist for coplanar prograde motion and  the librations' structure and width is similar to that of 1:2 resonance with the three critical inclination values $I_r=38^\circ$, $96^\circ$ and $138^\circ$ (Figure \ref{f4}).  Regardless of eccentricity, all resonant librations disappear for exactly coplanar retrograde motion $I_r=180^\circ$. 

Lastly, the 1:2 and 1:3 retrograde Neptune resonances cannot be examined by the current fourth order disturbing function as their second harmonics given by $\phi_6^{2:4}$ and $\phi_8^{2:6}$ respectively require expansions of order $N=6$ and $N=8$.\footnote{We used the property that for pure eccentricity retrograde resonances $k=p+q$ in  Section \ref{ss5.2}.} {We remind the reader that an expansion of order $N$ models all resonant angles $\phi^{p:q}_k=q\lambda-p\lambda^\prime-(q-p)\Omega -k \omega$  
with $|k|\leq N$ regardless of the values of $p$ and $q$ -- see, for instance, the expression of the direct part (\ref{RY}). }

\section{Conclusion}\label{section6}
\begin{figure*}
\begin{center}
{ 
\includegraphics[width=75mm]{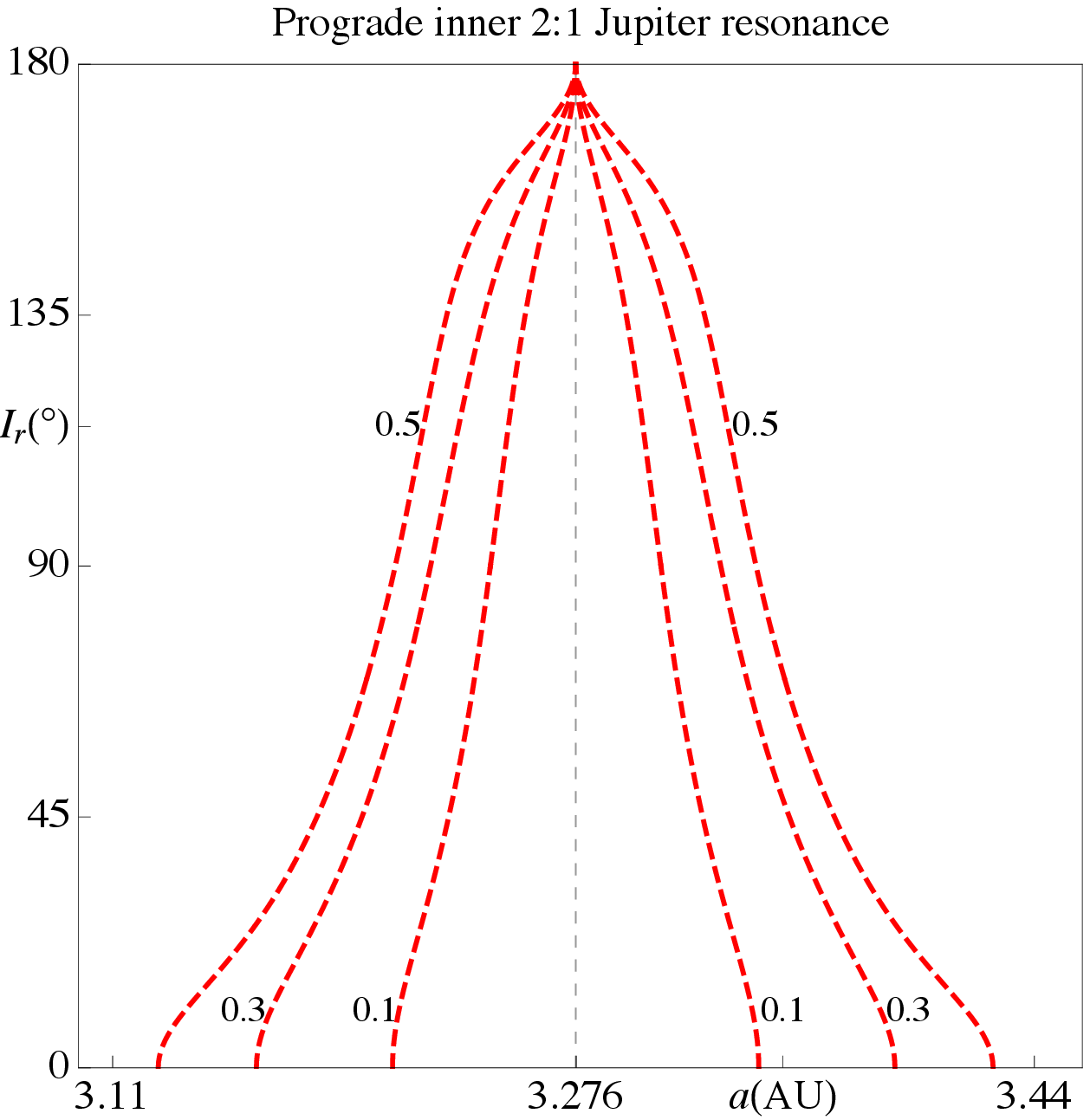}\hspace{10mm}\includegraphics[width=75mm]{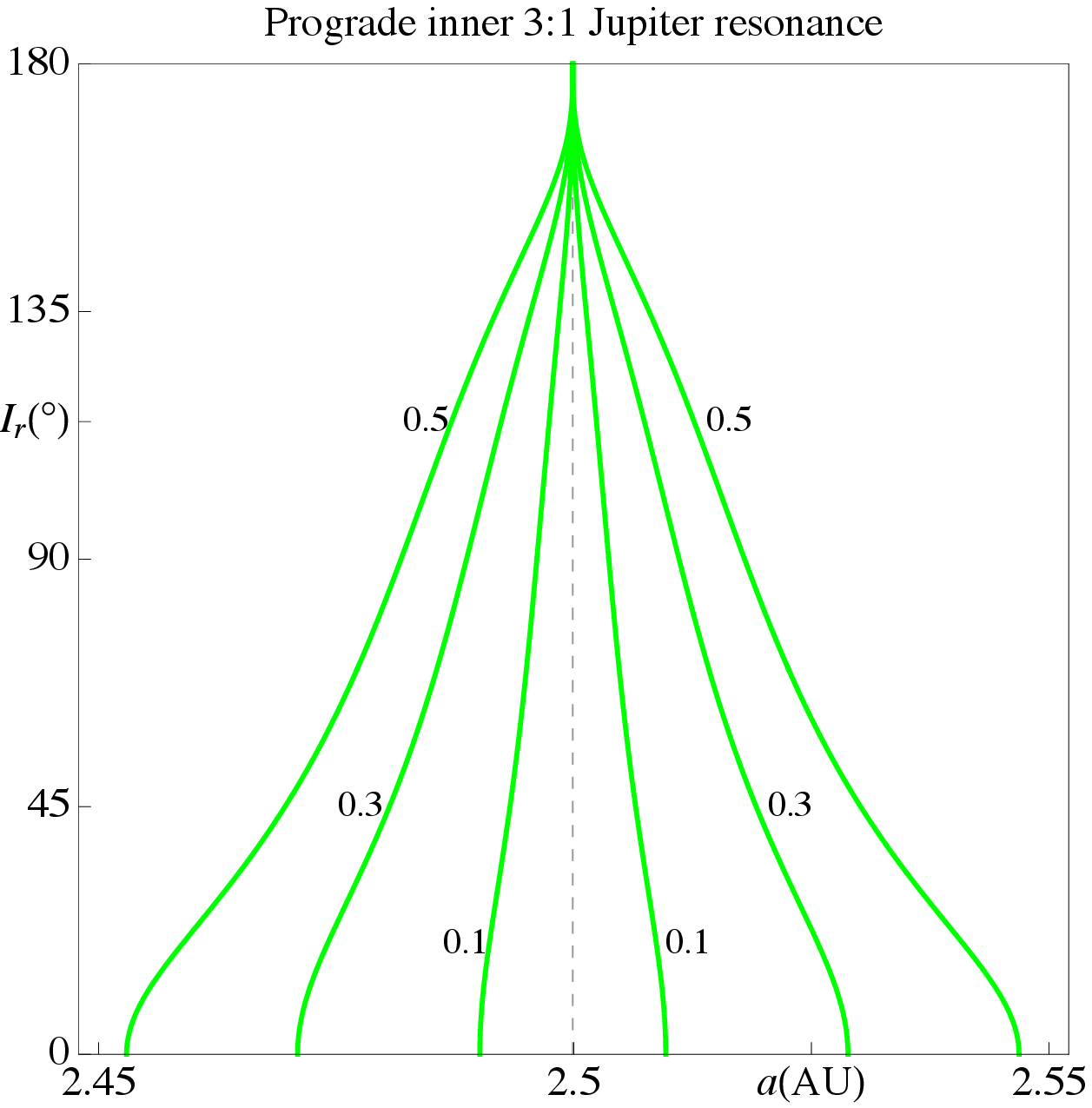}\\[5mm]
\includegraphics[width=75mm]{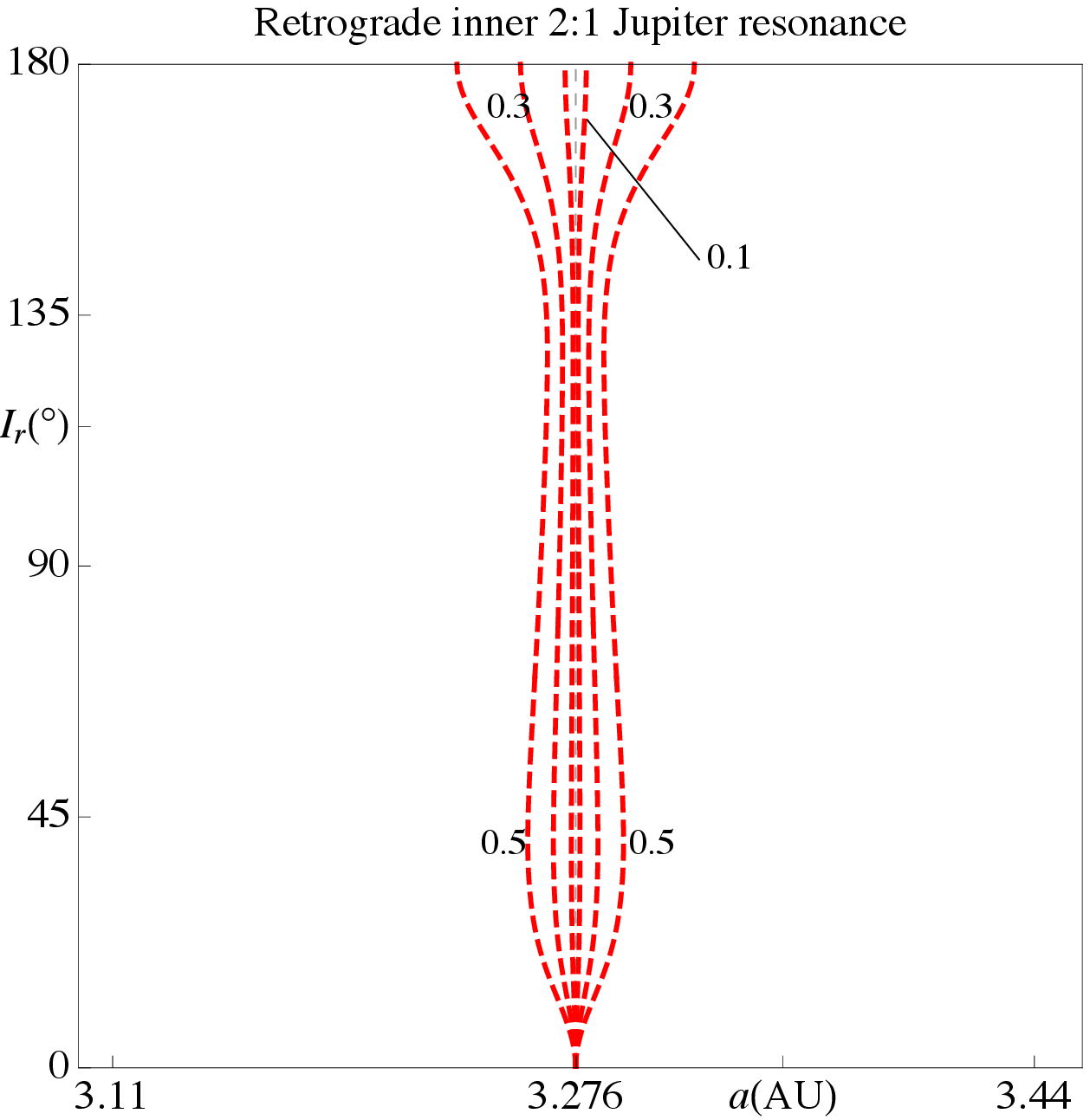}\hspace{10mm}\includegraphics[width=75mm]{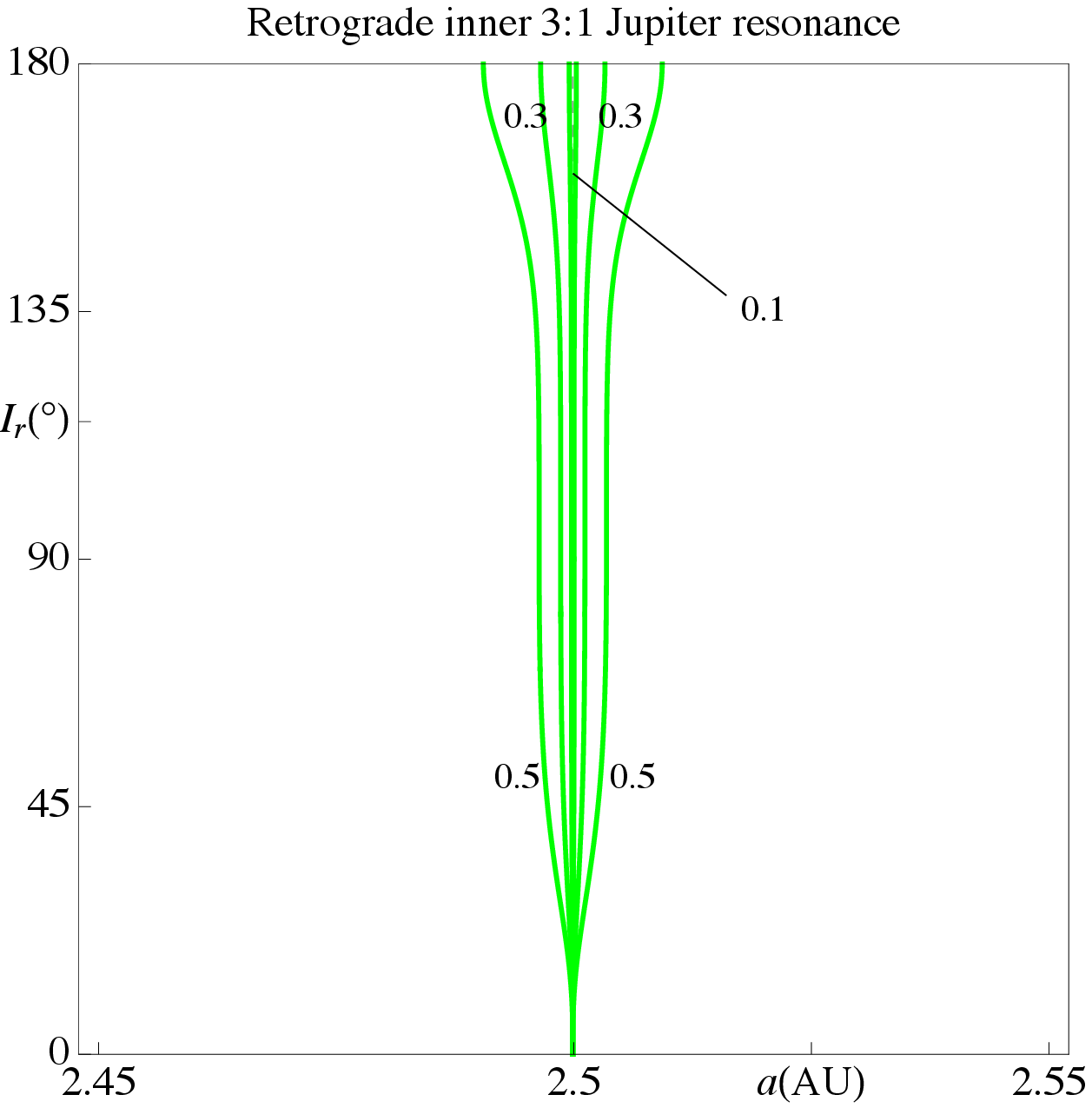}
}
\caption{Inner 2:1 and 3:1 Jupiter pure eccentricity resonance widths (\ref{reswidth0}) as functions of the asteroid's average inclination $I_r$ for three eccentricity values $e=0.1$, 0.3 and 0.5. Green solid lines indicate libration of the resonant argument around $180^\circ$ and red dashed lines around $0^\circ$. The dashed vertical line denotes the location of nominal resonance.  The number next to each curve is the corresponding eccentricity value.}\label{f2}
\end{center}
\end{figure*}

\begin{figure*}
\begin{center}
{ 
\includegraphics[width=75mm]{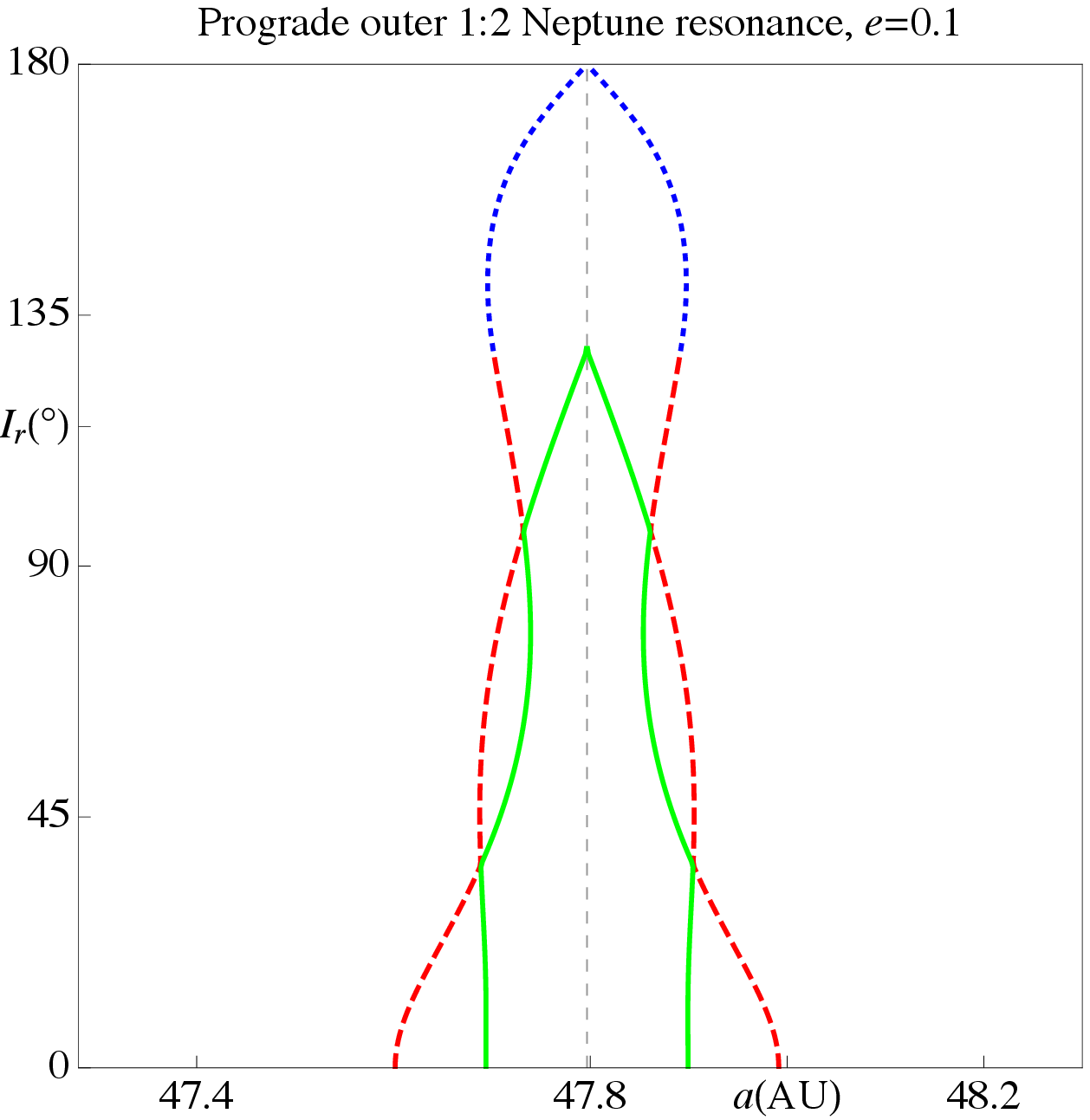}\hspace{10mm}\includegraphics[width=75mm]{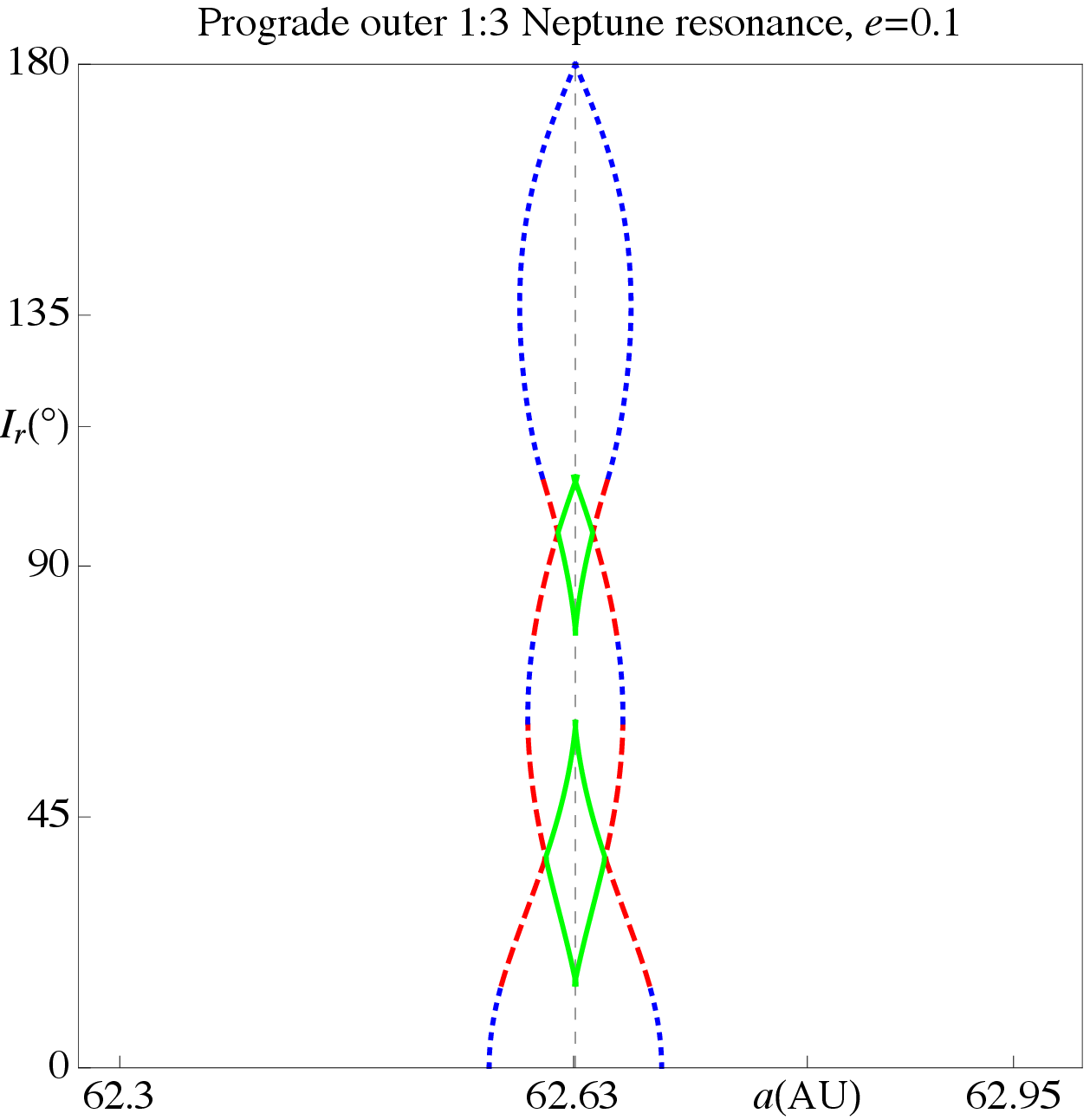}\\[5mm]
\includegraphics[width=75mm]{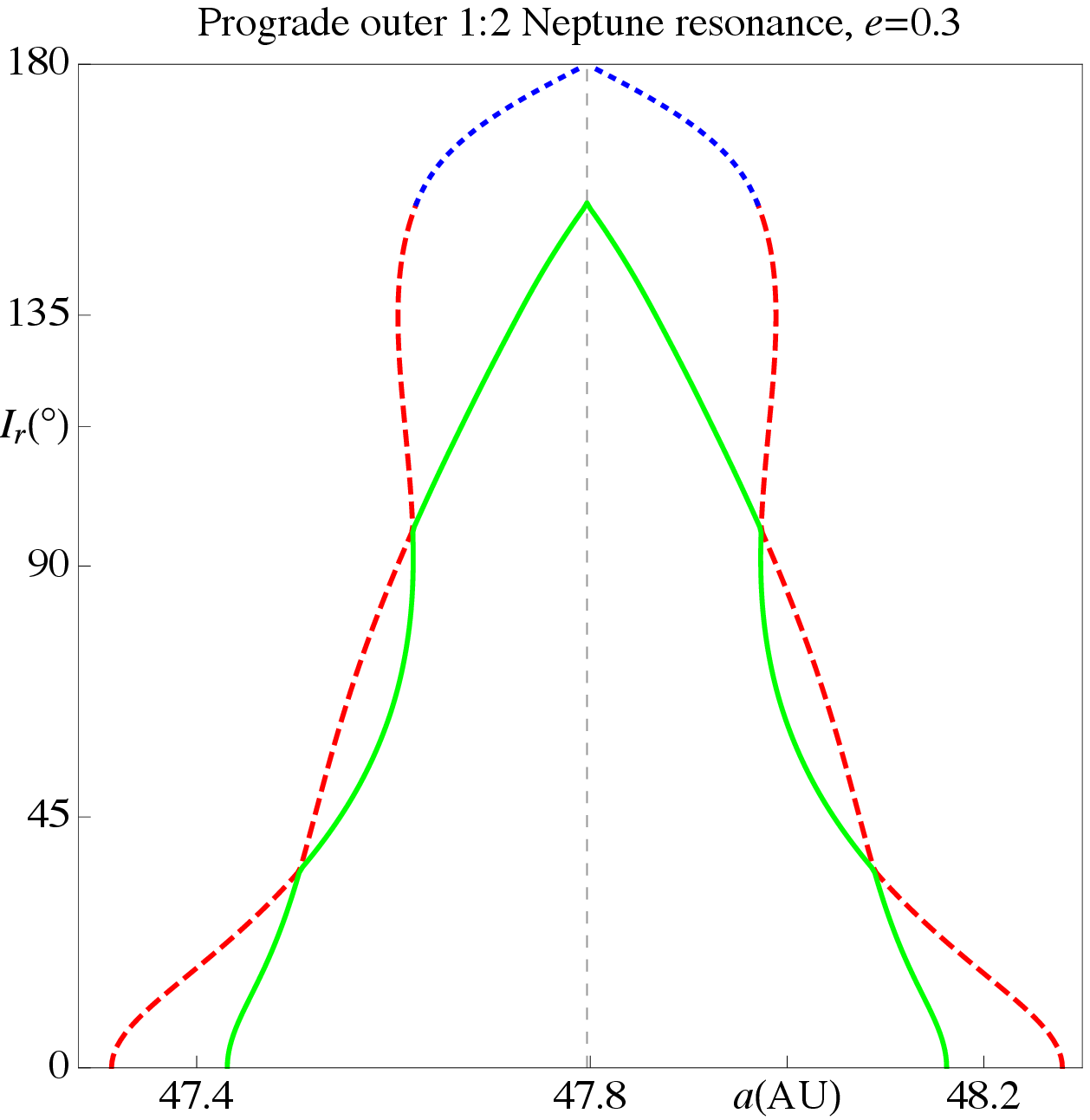}\hspace{10mm}\includegraphics[width=75mm]{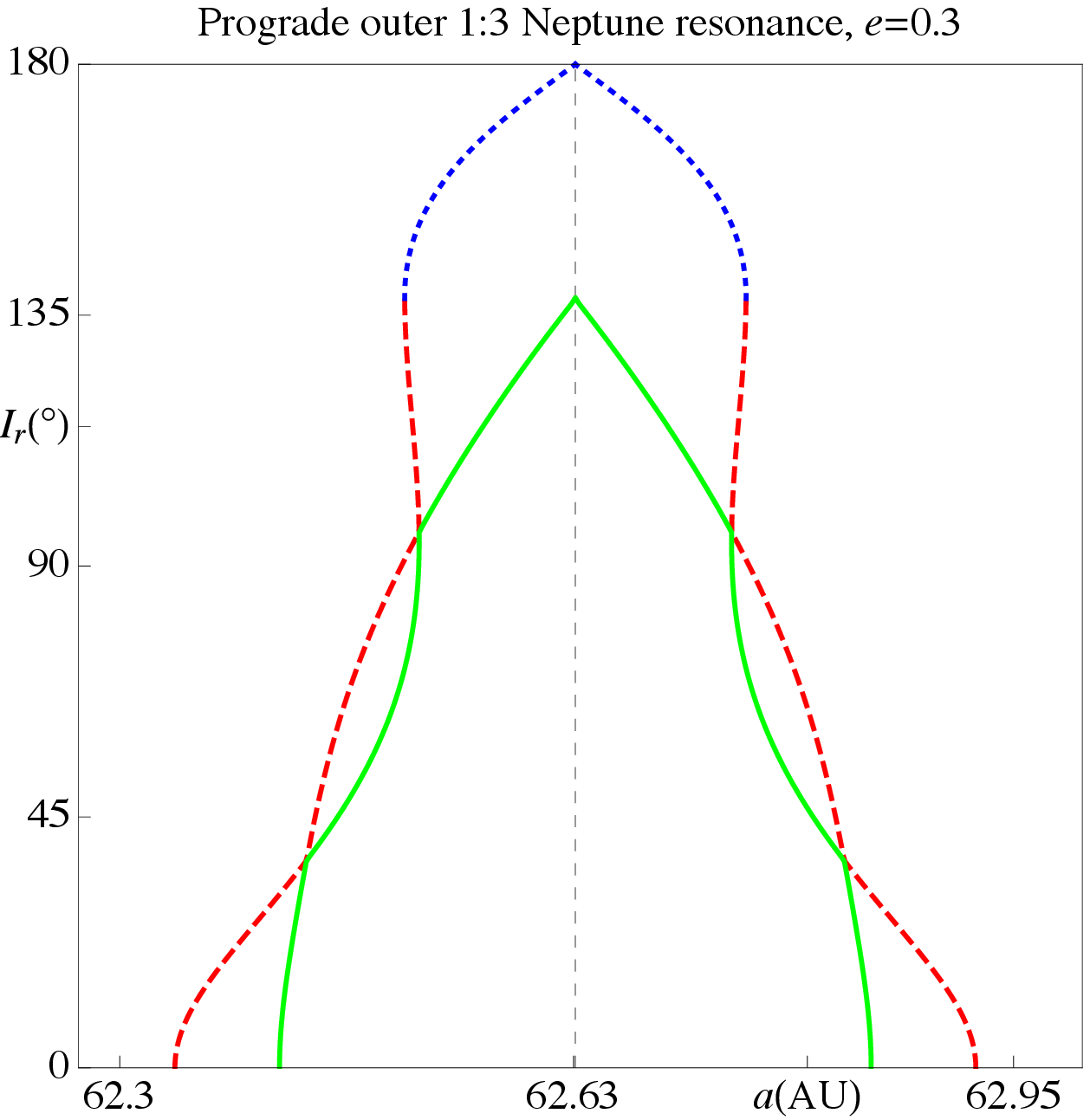}}
\caption{Outer 1:2 and 1:3 Neptune pure eccentricity resonance widths as functions of the asteroid's average inclination $I_r$ for two eccentricity values $e=0.1$ and 0.3. Solid green indicates the asymmetric resonance widths (\ref{reswidth2}), 
dashed red indicates the librations around $180^\circ$ encompassing both asymmetric centres (\ref{reswidth1}), and dotted blue indicates simple libration around $180^\circ$ when the second harmonic is weaker than a quarter of the first (\ref{reswidth0}). }\label{f3}
\end{center}
\end{figure*}

This work's main motivation of understanding the processes of resonance passage and capture at arbitrary inclination was inspired by our earlier numerical work on the subject \citep{NamouniMorais15,NamouniMorais17,NamouniMorais17c}, and encouraged by the realisation that the polar disturbing function of \citetalias{NamouniMorais17b} challenged the standard view on the force amplitudes associated with a resonant argument. We had therefore wondered whether the new properties of the polar disturbing function signalled a new general behaviour of disturbing functions that are based on the three-dimensional three-body problem. This work answers that query with the affirmative. We did so by providing the algorithm for the literal expansion of the gravitational interaction in the three-body problem with respect to an arbitrary inclination $I_r$ and giving explicitly its terms to fourth order in eccentricity $e$ and inclination $\sin (I-I_r)$.  We then showed how the new disturbing function reduces to the classical one by taking $I_r=0^\circ$. The properties of the classical disturbing function once-considered standard henceforth seem those of a degenerate expansion embodied by (i) the presence of the one-dimensional Laplace coefficients, (ii) the constraint that an expansion of order $N$ can only model resonances of order $N$ and (iii) the fact that only a pure eccentricity resonance of order $k$ has, to lowest order in eccentricity, a force amplitude $\propto e^{|k|}$ and no inclination dependence. These  once-thought fundamental properties are not true in general. When expansions are not carried out with respect to the two-dimensional problem, i.e. $I_r\neq 0^\circ$ and $I_r\neq 180^\circ$ in the first interpretation of the new disturbing function, or not carried out at all in the second interpretation of the new disturbing function with $s\equiv 0$, then the expansion of order $N$ may model any resonance order. More importantly, the force amplitudes of a resonant term {are} mostly independent of the resonance order giving, for instance the argument $\phi_1^{1:6}=6 \lambda-\lambda^\prime-5\Omega-\omega$, a force amplitude linear with respect to eccentricity that can be modelled by a series expansion of first order in eccentricity. To illustrate further the far-reaching consequences of the new disturbing function, we determined the inclination dependence of pure eccentricity resonances by measuring their widths with a new two-harmonics pendulum model. We found that contrary to the prediction of the classical disturbing function, those resonance depend strongly on inclination not only in terms of width but also in the nature of libration in resonance. For instance, we showed that below the threshold eccentricity of 0.13 known to exclude asymmetric librations for the pure eccentricity 1:3 outer resonance with Neptune, asymmetric librations do exist around inclinations of $45^\circ$ and $90^\circ$. That eccentricity threshold was derived numerically for the planar problem and thought to be valid in general as pure eccentricity resonances were believed to be independent of inclination. We now know that the 0.13 threshold is valid only for planar motion. Our preliminary numerical estimations (presented in a forthcoming article) using chaos indicators agree with the width and libration structures shown in Figures (\ref{f2}) and (\ref{f3}). {There are two natural extensions of the present work: firstly, 
the derivation of a disturbing function with an eccentric and inclined perturber. Our original motivation being the dynamical study  of small solar system bodies with arbitrary orbital inclinations, our approximation of a circular planet and our choice of its orbital plane as an inclination reference are quite sufficient. Indeed, we have shown in \citep{NamouniMorais15} that typical solar system planetary eccentricities  do not have a significant influence on mean motion resonance capture. Furthermore, as the inclinations of the solar system planets are small with respect the invariable Laplace surface \citep{tremaine09}, they will not influence significantly the dynamics of asteroids with large inclinations. 
The second extension consists of applying} the present disturbing function to derive analytically resonance capture probabilities at arbitrary inclination.

\newpage

\appendix 

\section{Two-dimensional Laplace coefficients with arbitrary reference inclination}\label{app}
In the following, we examine the properties of the new  two-dimensional Laplace coefficients that generalise the classical one-dimensional Laplace coefficients of the classical disturbing function of nearly coplanar orbits \citep{BC61}, and the two-dimensional Laplace coefficients of  the disturbing function of nearly polar orbits (\citetalias{NamouniMorais17b}). The new coefficients depend on the arbitrary reference inclination $I_r$ of the disturbing function and are defined as:
\begin{eqnarray}
&&b_{s}^{jk}(\alpha,I_r) = \frac{1}{\pi^2}\int_0^{2\pi}  \int_0^{2\pi} {\cos(j u+kv) \ du\, dv} \times\label{Lap0x}  \\
& &\ \ \ \ \ \ \ \ \ \ \ \ {[1+\alpha^2-2\alpha (\cos u \cos v-\sin u\sin v \cos I_r)]^{-s}}.\nonumber 
\end{eqnarray}
\subsection{Intrinsic properties}\label{A1}
We first recall the important property:
 \begin{equation}
 b_{s}^{jk}(\alpha,I_r)=0\ {\rm if\ }  j\pm k \ {\rm is\ odd}.
 \end{equation} 
The Laplace coefficients are defined for both $\alpha<1$ and $\alpha>1$ and satisfy the symmetry relation:
 \begin{equation}
b_{s}^{jk}(p,q,\alpha,I_r)=\alpha^{-2s}b_{s}^{jk}(p,q,\alpha^{-1},I_r). \label{Lap3} 
\end{equation}
This property is useful when seeking  a series expansion with respect to $\alpha>1$ (see Section \ref{A3}). 

The coefficients  $b_{s}^{jk}(\alpha,I_r)$  are not completely symmetric with respect to the angle integer coefficients $j$ and $k$ (as it is the case for polar motion i.e. $I_r=90^\circ$). Instead, we have the following identities that depend on the reference inclination $I_r$:
\begin{eqnarray}
 b_{s}^{(-j)(-k)}(\alpha,I_r) &=&b_{s}^{jk}(\alpha,I_r) =b^{kj}_s(\alpha,I_r),\label{Lap1}\\
 b_{s}^{(-j)k}(\alpha,I_r)&=&\ b_{s}^{j(-k)}(\alpha,I_r),\nonumber \\&=& b_{s}^{jk}(\alpha,I_r+180^\circ)\label{Lap2}.
 \end{eqnarray}
 In the last identity, the sum $I_r+180^\circ$ is defined modulo $180^\circ$.\footnote{One could equally write $I_r-180^\circ$ instead of $I_r+180^\circ$ in (\ref{Lap2}).} The previous properties naturally extend to the functions $A_{i,j,k,l}=\alpha^l D^l  b_{i+1/2}^{jk}$ (equation \ref{Aijkl}) as:
 \begin{eqnarray}
 A_{i,j,k,l}(\alpha,I_r)&=&0 \ {\rm if\ }  j\pm k \ {\rm is\ odd}, \label{ALap3}\\
A_{i,-j,-k,l}(\alpha,I_r)&=&A_{i,j,k,l}(\alpha,I_r), \nonumber \\&=&A_{i,k,j,l}(\alpha,I_r),\label{ALap1}\\
A_{i,-j,k,l}(\alpha,I_r)&=&A_{i,j,-k,l}(\alpha,I_r),\nonumber \\&=&  A_{i,j,k,l}(\alpha,I_r+180^\circ)\label{ALap2}.
\end{eqnarray}
 
 The derivatives of the Laplace coefficients also depend on the reference inclination as follows:
\begin{eqnarray}
D\, b_{s}^{jk}&=&\frac{s}{2} \left[(b_{s+1}^{(j+1)(k+1)}+b_{s+1}^{(j-1)(k-1)})(1+\cos I_r)+\right. \nonumber \\
&&\left.+(b_{s+1}^{(j+1)(k-1)}+b_{s+1}^{(j-1)(k+1)})(1-\cos I_r)\right]\label{derivB1}\\&&- 2 \alpha s b_{s+1}^{jk},\nonumber \\
D^n \, b_{s}^{jk}&=&\frac{s}{2} \left[\left(D^{n-1} \,b_{s+1}^{(j+1)(k+1)}+D^{n-1} \,b_{s+1}^{(j-1)(k-1)}\right)\times\right. \nonumber \\
&&(1+\cos I_r) \nonumber\\
&&+\left(D^{n-1} \,b_{s+1}^{(j-1)(k+1)}+D^{n-1} \,b_{s+1}^{(j+1)(k-1)}\right) \times \nonumber\\
&&\left.(1-\cos I_r)\right] - 2 \alpha s D^{n-1} \,b_{s+1}^{jk}\label{derivBn}\\ &&- 2(n-1)sD^{n-2}\,b_{s+1}^{jk}.\nonumber
\end{eqnarray}
where $D\equiv d/d\alpha$. 

\subsection{Relationship to the classical one-dimensional Laplace coefficients}\label{A2}
It is also useful to relate the new Laplace coefficients to those of the nearly coplanar disturbing function \citep{BC61}. Simple algebra shows  that $b_{s}^{jk}(\alpha,I_r=0^\circ)=0$ unless $k= j$ and  that $b_{s}^{jk}(\alpha,I_r=180^\circ)=0$ unless $k= -j$.  Furthermore:  
\begin{eqnarray}
b_{s}^{j j}(\alpha,0 ) &=& b_{s}^{j (-j)}(\alpha,180^\circ ) \label{Lap00} \\
&=& \frac{2}{\pi} \int_0^{2\pi} \frac{\cos(j u) \ du} {(1+\alpha^2-2\alpha \cos u)^{s}},\nonumber \\
&=& 2 b_s^j(\alpha)\label{Lap01}
\end{eqnarray}
where $ b_s^j(\alpha)$ is the classical one-dimensional Laplace coefficient. One may also check that the derivative relationships (\ref{derivB1},\ref{derivBn}) reduce to the known relations of the one-dimensional Laplace coefficients when $I_r=0^\circ$. The coplanar identity extends also to the functions $A_{i,j,k,l}$ as:
\begin{eqnarray} 
A_{i,j,k,l}(\alpha,I_r=0^\circ)&=&0 \ \ \ \  {\rm unless\  } k= j, \label{AIr0}\\ 
A_{i,j,k,l}(\alpha,I_r=180^\circ)&=& 0 \ \ \ \ {\rm unless\  } k= -j.\label{AIr180}
\end{eqnarray}
\subsection{Series expansion with respect to $\alpha$}\label{A3}
The numerical values of a given $b_{s}^{jk}$ coefficient and its derivatives may be obtained in two ways. One is to integrate numerically the expression (\ref{Lap0x}) and use the identities (\ref{derivB1},\ref{derivBn}) for its derivatives. Whereas the current computing power of a personal computer would be sufficient for small $j$ and $k$, the integral is highly oscillatory for larger integers making its evaluation time consuming. The other way is to seek a series expansion of the new two-dimensional Laplace coefficients with respect to the semimajor axis ratio $\alpha$ in the same way the classical one-dimensional Laplace coefficients are evaluated as a hypergeometric series (for details on the subject we refer the reader to \cite{BC61} p. 495). Series expansions with respect to the semimajor axis ratio suffer from the well-known feature of slow convergence when $\alpha \ll 1$ is not satisfied thus requiring large expansion orders.  Nonetheless the series method remains superior in terms {of} computation time to direct numerical integration of the full integral (\ref{Lap0x}) provided a sufficiently large expansion order is used. 

In order to determine the series expansion of the Laplace coefficient with respect to $\alpha$ to order $N_\alpha$, we may use the definition of Gegenbauer's polynomials (also known as ultra-spherical polynomials) as the coefficients of the series expansion of  $(1+\alpha^2-2\alpha x)^{-s}$ with respect to $\alpha$ \citep{AbraStegun}. This yields:
\begin{eqnarray}
b_{s}^{jk}&=& \frac{1}{\pi^2}\sum_{\nu=0}^{N_\alpha}
\alpha^\nu\int_0^{2\pi} \int_0^{2\pi}
\label{bGegen}
\\&&  \! \! \! \! \! \! \! \! \! \!\!C^s_\nu(\cos u \cos v-\sin u\sin v\cos I) \ \cos(j u + k v) \, du\, dv,\nonumber
\end{eqnarray}
where $C^s_\nu(x)$ are Gegenbauer's polynomials. They may be expressed from the hypergeometric function $_2F_1(-\nu,\nu+2s,s+1/2,z)=\nu! C^s_\nu(1-2z)/(2s)_\nu$  as the polynomial:
\begin{equation}
C^s_\nu(x)=\sum_{n=0}^{{\rm Floor}[\nu/2]} \frac{(-1)^n(s)_{(\nu-n)}}{n!(\nu-2n)!}(2x)^{\nu-2n}, \label{gege}
\end{equation}
where $(s)_\nu=s (s+1) (s+2) ... (s+\nu-1)$.
In the expression (\ref{bGegen}), the semimajor axis ratio $\alpha$ must be smaller than unity. The   Laplace coefficients for $\alpha>1$ are found using the relationship (\ref{Lap3}).
The integers $j$ and $k$ may also be restricted to $0\leq k \leq j$ as the Laplace coefficient of any other values may be brought to this case using the relationships (\ref{Lap1}) and (\ref{Lap2}). The argument of the Gegenbauer polynomial in the series (\ref{bGegen})  may be written as:
\begin{eqnarray}
x&=&\cos u \cos v-\sin u\sin v\cos I,\\
&=&[\eta_1\cos (u - v)+\eta_2 \cos(u+v)]/2,
\end{eqnarray}
where  $\eta_1=1-\cos I$, and $\eta_2=1+\cos I$. The Fourier coefficient of the Gegenbauer polynomials requires integrating the powers $x^m$ where $m=\nu-2n$. Expressing the cosines in $x$ in terms of the usual imaginary exponentials yields:
\begin{eqnarray}
x^m&=& 2^{-2m}\sum_{l=0}^{m}
{\left(\!\!\!\!{\scriptsize\begin{array}{c}m\\l\end{array}}\!\!\!\!\right)}  
\eta_1^{l} \eta_2^{m-l} \sum_{p=0}^l\sum_{q=0}^{m-l} 
{\left(\!\!\!{\scriptsize\begin{array}{c}l\\p\end{array}}\!\!\!\right)}
{\left(\!\!\!\!{\scriptsize\begin{array}{c}m-l\\q\end{array}}\!\!\!\right)} \nonumber \\
&&\ \ \ \ \ \ e^{i[2(q+p)-m]u}   e^{i[2(q-p)-m+2l]v}.
\end{eqnarray}
Multiplying $x^m$ by $\cos (ju +kv)=[e^{i(ju+kv)}+e^{-i(ju+kv)}]/2$, yields the sum of the two terms: 
\begin{eqnarray}
x^m_{jk+}&=& 2^{-2m-1}\sum_{l=0}^{m}
{\left(\!\!\!\!{\scriptsize\begin{array}{c}m\\l\end{array}}\!\!\!\!\right)}  
\eta_1^{l} \eta_2^{m-l} \sum_{p=0}^l\sum_{q=0}^{m-l} 
{\left(\!\!\!{\scriptsize\begin{array}{c}l\\p\end{array}}\!\!\!\right)}
{\left(\!\!\!\!{\scriptsize\begin{array}{c}m-l\\q\end{array}}\!\!\!\right)} \nonumber \\
&&\ \ \ \ \ \ e^{i[2(q+p)-m+j]u}   e^{i[2(q-p)-m+2l+k]v},\\
x^m_{jk-}&=& 2^{-2m-1}\sum_{l=0}^{m}
{\left(\!\!\!\!{\scriptsize\begin{array}{c}m\\l\end{array}}\!\!\!\!\right)}  
\eta_1^{l} \eta_2^{m-l} \sum_{p=0}^l\sum_{q=0}^{m-l} 
{\left(\!\!\!{\scriptsize\begin{array}{c}l\\p\end{array}}\!\!\!\right)}
{\left(\!\!\!\!{\scriptsize\begin{array}{c}m-l\\q\end{array}}\!\!\!\right)} \nonumber \\
&&\ \ \ \ \ \ e^{i[2(q+p)-m-j]u}   e^{i[2(q-p)-m+2l-k]v}.
\end{eqnarray}
Integrating $x^m_{jk+}$ and $x^m_{jk-}$  with respect to $u$ and $v$ between 0 and $2\pi$ requires that the powers of the exponentials be zero because otherwise the integrals vanish as they are $2\pi$-periodic. This gives the following conditions on $p$ and $q$:
\begin{eqnarray}
x^m_{jk+}&:& 2(q+p) =m-j, \ \  2(q-p)=m-k-2l,\nonumber\\
x^m_{jk-}&:& 2(q+p) =m+j, \ \  2(q-p)=m+k-2l.\nonumber
\end{eqnarray}
The first relation implies that $m$ must be of the same parity as $j$ (and therefore $k$) and that $m\geq j$. The integers $p$ and $q$ are determined uniquely as:
\begin{eqnarray}
x^m_{jk+}&:& \!\!\!p= \frac{2l-j+k}{4}, \ q= \frac{2(m-l)-(j+k)}{4},\\
x^m_{jk-}&:&\!\!\! p= \frac{2l+j-k}{4}, \ q= \frac{2(m-l)+(j+k)}{4}.
\end{eqnarray}
The first and third equations require that $l$ have the same parity as $(j-k)/2$. The second and fourth require that $m-l$ have the same parity as $(j+k)/2$ (which is that of $(j-k)/2$ if $k$ and $j$ are even and its opposite if the integers are odd). This last condition is redundant as when combined with the first, it shows that $m$ has the same parity as $j$ and $k$. For the integers $p$ and $q$ to exist they must satisfy $0\leq p\leq l$ and $0 \leq q \leq m-l$. This translates into the following relations for $x^m_{jk+}$:
\begin{eqnarray}
 p\geq 0 \ {\rm implies}&&\!\!\!\!\!\!\!  l\geq \frac{j-k}{2}, \label{cond1p}\\
 p\leq  l \ {\rm implies}&&\!\!\!\!\!\!\!  l \geq \frac{k-j}{2}, \\
&&\!\!\!\!\!\!\! {\rm  trivial \ condition\ as\ } j\geq k,\nonumber \\
 q\geq 0 \ {\rm implies}&&\!\!\!\!\!\!\!   l \leq m-\frac{j+k}{2}, \label{cond3p}\\
 q\leq m-l  \ {\rm implies}&&\!\!\!\!\!\!\!   l \leq m+\frac{j+k}{2},\\
 && \!\!\!\!\!\!\! {\rm trivial \ condition\ as\ } l\leq m.\nonumber
\end{eqnarray}
Conditions (\ref{cond1p},\ref{cond3p}) determine the range of  the $l$-sum as  $(j-k)/2\leq l \leq m-(j+k)/2$ with  $l$ of the same parity as $(j-k)/2$. For the second term  $x^m_{jk-}$, the requirements yield exactly the same summation range for the $l$-sum. Calling $y^m_{jk\pm}=\pi^{-2}\int_{0}^{2\pi}\int_{0}^{2\pi} x^m_{jk\pm}du \ dv$, simple algebra shows that $y^m_{jk+}=y^m_{jk-}$ and:
\begin{eqnarray}
y^m_{jk\pm}&=&\!\!\!\!\! \frac{m!}{2^{2m-1}}\!\!\!\!\!\!\! \!\!\!\!\!\!\! \!\!\!\!\!\!\! \sum_{
{\scriptsize\begin{array}{c}
\\
 \frac{j-k}{2}\leq l \leq m-\frac{j+k}{2}\\
l \ {\rm same\ parity\ as }\ \frac{j-k}{2}\ \end{array}}}\!\!\!\!\! \!\!\!\!\!\!\! \frac{ \eta_1^{l} \eta_2^{m-l}}{[(2l-j+k)/4]![(2l+j-k)/4]!}\nonumber\\
&\times&\!\!\!\!\!\frac{1}{[(2m-2l-k-j)/4]![(2m-2l+k+j)/4]!}. \nonumber\label{yp0}
\end{eqnarray}
The summation over $l$ can be simplified by removing the parity condition through the transformation $l\rightarrow 2 l +(j-k)/2$ that  gives a more compact form of $y^m_{jk\pm}$ as:
\begin{eqnarray}
y^m_{jk\pm}&=&\!\!\!\!\! \frac{m!\eta_1^{\frac{j-k}{2}} \eta_2^{\frac{k-j}{2}}}{2^{2m-1}}\sum_{l=0}^{{\rm Floor}[(m-j)/2]}\!\!\!\!\! \!\!\!\!\!\!\! \frac{ \eta_1^{2l} \eta_2^{m-2l}}{l![(2l+j-k)/2]!}\nonumber\\
&\times&\!\!\!\!\!\frac{1}{[(m-2l-j)/2]![(m-2l+k)/2]!}. \label{yp}
\end{eqnarray}

The condition that  $m$ must have the same parity as $j$ and satisfy $m\geq j$ restricts the Gegenbauer polynomials (\ref{gege}) to $0 \leq n \leq {\rm Floor}[(\nu-j)/2]$ and the full two-dimensional Laplace coefficient series  (\ref{bGegen}) to $\nu\geq j$ and $\nu$ of the same parity as $j$. By substituting twice the sum (\ref{yp})  into the expression of the Gegenbauer polynomials (\ref{gege}), and applying the index change $\nu\rightarrow 2 \nu+j$ the  series expansion of the two-dimensional Laplace coefficient with arbitrary inclination with respect to the semimajor axis ratio $\alpha$  to order $N_\alpha$ (\ref{bGegen}) may be written as:
\begin{eqnarray}
b_{s}^{jk}&=&  \!\!\!\!\!  \frac{ (1-\cos I_r)^\frac{j-k}{2} (1+\cos I_r)^\frac{j+k}{2}\, \alpha^j}{2^{j-2}} \times \nonumber\\ 
&&\!\!\!\!\! \sum_{\nu=0}^{{\rm Floor}[(N_\alpha-j)/2]}
\alpha^{2\nu}\sum_{n=0}^{\nu} \frac{(-1)^n(s)_{(2\nu+j-n)}}{n!2^{2(\nu-n)}}\times \nonumber \\ 
&&\!\!\!\!\! \sum_{l=0}^{\nu-n} \frac{ (1-\cos I_r)^{2l} (1+\cos I_r)^{2(\nu-n-l)}}{l![l+(j-k)/2]!}\times \nonumber\\
&&\ \ \ \frac{1}{(\nu-n-l)![\nu-n-l+(j+k)/2]!}. \label{bsjk}
\end{eqnarray}
We remind the reader that  expression (\ref{bsjk}) assumes $\alpha<1$ and $0
\leq k\leq j$. For different parameters, the symmetries of the Laplace coefficients (\ref{Lap3}, \ref{Lap1}, \ref{Lap2}) may be used to get back to the last two conditions.

For $I_r=0^\circ$ and $j=k$ (as required by \ref{Lap00}),  all terms in the $l$-sum vanish except $l=0$ thus simplifying  the general expression (\ref{bsjk}) to:
\begin{eqnarray}
b_{s}^{jj}(I_r=0^\circ) \!\!\!\!\!&=&   \!\!\!\!\!4 \alpha^j \!\!\!\!\!\!\!\sum_{\nu=0}^{{\rm Floor}[(N_\alpha-j)/2]} \!\!\!\!\!
\alpha^{2\nu}\sum_{n=0}^\nu\frac{(-1)^n(s)_{(2\nu-n+j)}}{n!(\nu-n)!(\nu-n+j)!}. \nonumber\\
&& 
\end{eqnarray}
which equals twice the classical series of the one-dimensional Laplace coefficient (\ref{Lap01}) \citep{BC61}. 

The series expansion of the Laplace coefficient (\ref{bsjk})  is used in Section \ref{section5} to determine resonance widths that utilise  $b_s^{jj}$ for $0^\circ\leq I_r\leq 180^\circ$. Comparison with the direct numerical integration of the expression (\ref{Lap0x}) shows that for the 2:1 and hence 1:2 resonance locations, an expansion order of $N_\alpha=20$ is required to reach a $10^{-6}$  relative error whereas for the 3:1 and hence the 1:3 resonance locations, the relative error is $10^{-9}$ with the same expansion order.

\section{Fourth order series expansion with arbitrary inclination} \label{appB}
In this Appendix, we produce the explicit force coefficients of the disturbing function with arbitrary reference inclination. The first five Tables contain the force coefficients of the cosine terms $\phi_k^{p:q}=\phi-k\omega$ where $\phi=q\lambda-p\lambda^\prime -(q-p)\Omega$. The indirect part of the disturbing function is given in Table \ref{t6} and the secular potential in Table \ref{t7}. The force coefficients use the quantities $A_{i, p, q, l}=\alpha^l D^l  b_{i+1/2}^{jk}$ that appear in the literal expansion (equation \ref{Aijkl})  and whose properties are further discussed in Appendix \ref{app}.
\begin{table}
  \caption{Force coefficients $c^0_{mn}(p,q,\alpha,I_r)$ of the term $e^m s^n  \cos\phi$. }\label{t1}
   \begin{tabular}{cl}
    \hline
$c_{00}^0 $&$\frac{1}{2}A_{0, p, q, 0},$\\
$c_{01}^0 $&$ -\frac{\alpha}{8} (A_{1, p-1, q-1, 0} - A_{1, p-1, q+1, 0} -  A_{1, p+1, q-1, 0} $\nonumber\\&$+ A_{1, p+1, q+1, 0}) \sin I_r,$\\
$c_{20}^0 $&$ \frac{1}{8} (-4 q^2 A_{0, p, q, 0} + 2 A_{0, p, q, 1} +  A_{0, p, q, 2}),$\\
$c_{02}^0 $&$\frac{3\alpha^2}{64} (A_{2, p-2, q-2, 0} - 2 A_{2, p-2, q, 0} + 
A_{2, p-2, q+2, 0}$\nonumber\\&$ - 2 A_{2, p, q-2, 0}+ 4 A_{2, p, q, 0} - 
2 A_{2, p, q+2, 0} $\nonumber\\&$+ A_{2, p+2, q-2, 0} - 
2 A_{2, p+2, q, 0} + A_{2, p+2, q+2, 0})\sin^2I_r$\\
& $-\frac{\alpha}{16}  (A_{1, p-1, q-1, 0} - A_{1, p-1, q+1, 0} - 
   A_{1, p+1, q-1, 0} $\\&$+ A_{1, p+1, q+1, 0})\cos I_r$\\
$c_{21}^0 $&$\frac{\alpha}{32} [( 4 q^2-2) (A_{1, p-1, q-1, 0}-A_{1, p-1, q+1, 0}$\nonumber\\&$-A_{1, p+1, q-1, 0}+A_{1, p+1, q+1, 0} ) - 4 (A_{1, p-1, q-1, 1}$\nonumber\\&$+A_{1, p-1, q+1, 1}+A_{1, p+1, q-1, 1} -A_{1, p+1, q+1, 1}) $\nonumber\\&$- A_{1, p-1, q-1, 2} + A_{1, p-1, q+1, 2} + A_{1, p+1, q-1, 2}   $\nonumber\\&$- A_{1, p+1, q+1, 2}]\sin I_r$\\
$c_{03}^0 $&$-\frac{5\alpha^3}{256} [ 3 (A_{3, p-3, q+1, 0}  - 
    A_{3, p-3, q-1, 0}  -  A_{3, p-1, q-3, 0}$\nonumber\\&$ + 
    A_{3, p-1, q+3, 0} +  A_{3, p+1, q-3, 0}  - 
    A_{3, p+1, q+3, 0} $\nonumber\\&$ + 
    A_{3, p+3, q-1, 0} - A_{3, p+3, q+1, 0}) +
   9( A_{3, p-1, q-1, 0} $\nonumber\\&$-  A_{3, p-1, q+1, 0} - 
    A_{3, p+1, q-1, 0} +  A_{3, p+1, q+1, 0}) $\nonumber\\&$- A_{3, p+3, q-3, 0}- 
   A_{3, p-3, q+3, 0}+    A_{3, p-3, q-3, 0}$\nonumber\\&$+A_{3, p+3, q+3, 0}] \sin^3I_r
    +\frac{3\alpha^2}{64} [A_{2, p-2, q-2, 0}$\\ &$ + A_{2, p+2, q-2, 0}+ A_{2, p+2, q+2, 0} + 
   A_{2, p-2, q+2, 0} $\\&$ +  2(2A_{2, p, q, 0}-  A_{2, p-2, q, 0} - 
    A_{2, p, q+2, 0} $\\&$ -   A_{2, p+2, q, 0} -  A_{2, p, q-2, 0})] \cos I_r \sin I_r$\\
    $c_{40}^0 $&$\frac{ 1} {128}[q^2 (16 q^2-9) A_{0, p, q, 0} - 8 q^2 (A_{0, p, q, 1}+ 
    A_{0, p, q, 2})   $\nonumber\\&$+ 4 A_{0, p, q, 3} + A_{0, p, q, 4}]$\\
$c_{04}^0 $&$\frac{ 3\alpha^2} {32768}[64 (A_{2, p-2, q-2, 0} - 
    2A_{2, p-2, q, 0} $\\&$+ A_{2, p-2, q+2, 0} - 
    2 A_{2, p, q-2, 0} + 4 A_{2, p, q, 0} $\\&$- 
    2 A_{2, p, q+2, 0} + A_{2, p+2, q-2, 0} - 
    2 A_{2, p+2, q, 0} $\\&$+ A_{2, p+2, q+2, 0}) + 
    35 \alpha^2  (A_{4, p-4, q-4, 0} $\\&$+ 
        A_{4, p+4, q-4, 0} +
        A_{4, p-4, q+4, 0} + 
        A_{4, p+4, q+4, 0} $\\&$ - 
    4   (A_{4, p+2, q-4, 0}  +
       A_{4, p+4, q+2, 0}+
       A_{4, p+2, q+4, 0} $\\&$+ 
       A_{4, p+4, q-2, 0} +
       A_{4, p-2, q+4, 0}+
       A_{4, p-4, q-2, 0}  $\\&$+ 
       A_{4, p-4, q+2, 0} +
       A_{4, p-2, q-4, 0}) + 
    6 (  A_{4, p-4, q, 0}$\\&$+ 
       A_{4, p, q-4, 0}+ 
       A_{4, p, q+4, 0}+ 
       A_{4, p+4, q, 0})$\\&$+ 
    16  ( A_{4, p-2, q-2, 0} +
      A_{4, p-2, q+2, 0} + 
      A_{4, p+2, q-2, 0}$\\&$ + 
      A_{4, p+2, q+2, 0}) - 
    24   (A_{4, p, q-2, 0} +
     A_{4, p-2, q, 0}$\\&$ +
      A_{4, p, q+2, 0}+
       A_{4, p+2, q, 0} )  + 
    36   A_{4, p, q, 0}   )]  $\\&$
    +\frac{\alpha}{2048}  [32 (A_{1, p+1, q-1, 0}- A_{1, p-1, q-1, 0}$\\&$ + 
     A_{1, p-1, q+1, 0}  - 
     A_{1, p+1, q+1, 0}) + 
    15 \alpha^2 (  
        A_{3, p+3, q-3, 0}$\\&$-A_{3, p-3, q-3, 0} - 
        A_{3, p+3, q+3, 0}+ 
        A_{3, p-3, q+3, 0}$\\&$+ 
    3   (A_{3, p-3, q-1, 0} - 
       A_{3, p-3, q+1, 0}  + 
       A_{3, p-1, q-3, 0} $\\&$- 
       A_{3, p-1, q+3, 0} - 
       A_{3, p+1, q-3, 0}+ 
       A_{3, p+1, q+3, 0}$\\&$ - 
       A_{3, p+3, q-1, 0} + 
       A_{3, p+3, q+1, 0}) + 
    9  ( A_{3, p+1, q-1, 0} $\\&$- 
       A_{3, p+1, q+1, 0}  - 
       A_{3, p-1, q-1, 0} + 
       A_{3, p-1, q+1, 0}) )]\cos I_r$\\
       
               \hline
  \end{tabular}
 \end{table}

     \setcounter{table}{0}
     
\begin{table}
  \caption{Continued. }
   \begin{tabular}{cl}
    \hline\\
  
       &$+  
    \frac{\alpha^2}{8192} [48 (A_{2, p-2, q-2, 0} - 
    2 A_{2, p-2, q, 0}$\\&$ +  A_{2, p-2, q+2, 0} - 
    2 A_{2, p, q-2, 0} + 4 A_{2, p, q, 0} $\\&$- 
    2 A_{2, p, q+2, 0} +  A_{2, p+2, q-2, 0} - 
    2A_{2, p+2, q, 0} $\\&$+  A_{2, p+2, q+2, 0}) +
    35 \alpha^2 ( 
    4   (A_{4, p-4, q-2, 0}$\\&$+ 
      A_{4, p-2, q+4, 0}+ 
       A_{4, p-4, q+2, 0}+ 
       A_{4, p-2, q-4, 0} $\\&$ + 
       A_{4, p+2, q-4, 0} + 
       A_{4, p+2, q+4, 0}+ 
       A_{4, p+4, q-2, 0}$\\&$+ 
       A_{4, p+4, q+2, 0} )-A_{4, p-4, q-4, 0}  - 
       A_{4, p+4, q-4, 0} $\\&$ - 
       A_{4, p-4, q+4, 0}  - 
       A_{4, p+4, q+4, 0} - 
    6(   A_{4, p, q-4, 0}$\\&$ + 
      A_{4, p-4, q, 0}  +
       A_{4, p, q+4, 0} +
       A_{4, p+4, q, 0})$\\&$ - 
    16   (A_{4, p+2, q-2, 0} +
      A_{4, p+2, q+2, 0}+ 
      A_{4, p-2, q-2, 0} $\\&$+ 
      A_{4, p-2, q+2, 0})+ 
    24   (A_{4, p, q-2, 0} + 
       A_{4, p, q+2, 0}$\\&$+ 
       A_{4, p+2, q, 0}+ 
      A_{4, p-2, q, 0} )- 
    36   A_{4, p, q, 0}   )] \cos 2I_r  $\\
   &$ 
    +\frac{15\alpha^3}{2048}  [A_{3, p-3, q-3, 0}- A_{3, p+3, q-3, 0}$\\&$ - 
    A_{3, p-3, q+3, 0}+ 
    A_{3, p+3, q+3, 0} + 3 (A_{3, p-3, q+1, 0} $\\&$- 
    A_{3, p-3, q-1, 0}  -  A_{3, p-1, q-3, 0} + 
     A_{3, p-1, q+3, 0} $\\&$+  A_{3, p+1, q-3, 0}  - 
     A_{3, p+1, q+3, 0} + 
     A_{3, p+3, q-1, 0} $\\&$-  A_{3, p+3, q+1, 0})+ 9 (A_{3, p+1, q+1, 0}- 
     A_{3, p+1, q-1, 0}$\\&$ + 
     A_{3, p-1, q-1, 0} -  A_{3, p-1, q+1, 0}) ] \cos 3 I_r   $\\
    &$
    +\frac{35 \alpha^4}{32768}  (A_{4, p-4, q-4, 0} + A_{4, p-4, q+4, 0}$\\&$+ 
    A_{4, p+4, q+4, 0} + 
    A_{4, p+4, q-4, 0} - 
    4( A_{4, p-4, q-2, 0} $\\&$ + 
     A_{4, p-4, q+2, 0} +
     A_{4, p-2, q+4, 0}  +  A_{4, p+2, q+4, 0}$\\&$+A_{4, p+4, q-2, 0}+ A_{4, p+4, q+2, 0}+ 
     A_{4, p-2, q-4, 0} $\\&$ +  A_{4, p+2, q-4, 0})  + 6 (A_{4, p, q-4, 0}+  A_{4, p-4, q, 0}$\\&$+ 
     A_{4, p, q+4, 0}+ 
     A_{4, p+4, q, 0}) + 16 (A_{4, p-2, q-2, 0} $\\&$ +  A_{4, p-2, q+2, 0} + 
     A_{4, p+2, q+2, 0}+ 
     A_{4, p+2, q-2, 0} )$\\&$
- 24 (A_{4, p+2, q, 0}   +
    A_{4, p-2, q, 0} + A_{4, p, q-2, 0}$\\&$+A_{4, p, q+2, 0})  + 36 A_{4, p, q, 0}  ] \cos 4 I_r
    $\\
   $c_{22}^0 $&$-\frac{3\alpha^2}{256} [2( 2 q^2-3) (A_{2, p-2, q-2, 0}-2 A_{2, p-2, q, 0}$\nonumber\\&$+ A_{2, p-2, q+2, 0} - 2 A_{2, p, q-2, 0}
  - 2 A_{2, p, q+2, 0} $\nonumber\\&$+    4A_{2, p, q, 0}-2 A_{2, p+2, q, 0} +A_{2, p+2, q-2, 0}$\nonumber\\&$+A_{2, p+2, q+2, 0}) - 
   6 A_{2, p-2, q-2, 1} - A_{2, p-2, q-2, 2}  $\nonumber\\&$+ 
   12 A_{2, p-2, q, 1} + 2 A_{2, p-2, q, 2}  - 
   6 A_{2, p-2, q+2, 1} $\nonumber\\&$- A_{2, p-2, q+2, 2}  + 
   12 A_{2, p, q-2, 1} + 2 A_{2, p, q-2, 2}  $\nonumber\\&$- 24 A_{2, p, q, 1} - 4 A_{2, p, q, 2} + 
   12 A_{2, p, q+2, 1} $\nonumber\\&$+ 2 A_{2, p, q+2, 2}  - 
   6 A_{2, p+2, q-2, 1} - A_{2, p+2, q-2, 2} $\nonumber\\&$ + 
   12 A_{2, p+2, q, 1} + 2 A_{2, p+2, q, 2} - 
   6 A_{2, p+2, q+2, 1}$\nonumber\\&$ - A_{2, p+2, q+2, 2}]\sin^2I_r+
    \frac{\alpha}{64}  [2( 2q^2-1) (A_{1, p-1, q-1, 0}$\\&$-A_{1, p-1, q+1, 0}-A_{1, p+1, q-1, 0}+A_{1, p+1, q+1, 0}) $\\&$+ 
   4 (A_{1, p-1, q+1, 1} 
   -  A_{1, p-1, q-1, 1}   + 
    A_{1, p+1, q-1, 1} $\\&$ - 
    A_{1, p+1, q+1, 1})+ A_{1, p+1, q-1, 2}  + A_{1, p-1, q+1, 2}$\\&$- A_{1, p-1, q-1, 2} - A_{1, p+1, q+1, 2}] \cos I_r$\\
 \hline
  \end{tabular}
 \end{table}

\begin{table}\label{t2}
  \caption{Force coefficients $c^{1}_{mn}(p,q,\alpha,I_r)$ of the term $e^ms^n \cos(\phi-\omega)$. }\label{t2}
   \begin{tabular}{cl}
    \hline
$c_{10}^{1} $&$-\frac{1}{4} [2 (1 - q) A_{0, p, q-1, 0} + A_{0, p, q-1, 1}],$\\ 
$c_{11}^{1} $&$ -\frac{\alpha}{16}  [( 2 q-3) (A_{1, p-1, q-2, 0} - A_{1, p-1, q, 0}+ A_{1, p+1, q, 0}$\nonumber\\&$- A_{1, p+1, q-2, 0}) - A_{1, p-1, q-2, 1}  + A_{1, p-1, q, 1} $\nonumber\\&$ +  A_{1, p+1, q-2, 1}  - A_{1, p+1, q, 1}]\sin I_r,$\\
$c_{12}^{1} $&$\frac{3\alpha^2}{128} [2 (2-q) (A_{2, p-2, q-3, 0}-2A_{2, p-2, q-1, 0} $\nonumber\\&$+A_{2, p-2, q+1, 0} 
                              -2A_{2, p, q-3, 0}+4A_{2, p, q-1, 0}$\nonumber\\&$+ A_{2, p+2, q-3, 0}-2A_{2, p, q+1, 0}-2 A_{2, p+2, q-1, 0}$\nonumber\\&$+A_{2, p+2, q+1, 0} ) + 
   A_{2, p-2, q-3, 1} 
    - 2 A_{2, p-2, q-1, 1} $\nonumber\\&$
    + A_{2, p-2, q+1, 1} 
   - 2 A_{2, p, q-3, 1} 
     + 4 A_{2, p, q-1, 1}$\nonumber\\&$
       - 2 A_{2, p, q+1, 1} 
       + A_{2, p+2, q-3, 1} 
        - 2 A_{2, p+2, q-1, 1} $\nonumber\\&$
       + A_{2, p+2, q+1, 1}]\sin^2I_r+      
       \frac{\alpha}{32}[(3 - 2 q) (A_{1, p-1, q-2, 0} $\\&$- A_{1, p-1, q, 0}-A_{1, p+1, q-2, 0}+A_{1, p+1, q, 0})$\\&$
  + A_{1, p-1, q-2, 1} - A_{1, p-1, q, 1}  - 
   A_{1, p+1, q-2, 1}   $\\&$+ A_{1, p+1, q, 1}] \cos I_r $\\   
$c_{30}^1 $&$\frac{1}{32} [2 q (7 q - 4 q^2-3) A_{0, p, q-1, 0} + 
   q (4 q-1) A_{0, p, q-1, 1} $\nonumber\\&$+(2q- 4) A_{0, p, q-1, 2} 
    - A_{0, p, q-1, 3}]$\\
 $c_{13}^1 $&$
 \frac{5\alpha^3}{2048}[(2 q-5) (A_{3, p-3, q-4, 0}-3 A_{3, p-3, q-2, 0}$\\&$ + 3 A_{3, p-3, q, 0}-  A_{3, p-3, q+2, 0} - 
   3 A_{3, p-1, q-4, 0} $\\&$+ 9A_{3, p-1, q-2, 0}- 
    9 A_{3, p-1, q, 0} + 3 A_{3, p-1, q+2, 0}$\\&$+ 3A_{3, p+1, q-4, 0}-  9A_{3, p+1, q-2, 0}+ 9A_{3, p+1, q, 0}$\\&$
     -  3A_{3, p+1, q+2, 0}-
   A_{3, p+3, q-4, 0}- 3 A_{3, p+3, q, 0} $\\&$+3A_{3, p+3, q-2, 0} +   A_{3, p+3, q+2, 0})  
   + 3 (A_{3, p-3, q-2, 1} $\\&$  - 
   A_{3, p-3, q, 1}   + 
    A_{3, p-1, q-4, 1} - A_{3, p-1, q+2, 1}$\\&$  - 
    A_{3, p+1, q-4, 1} +  A_{3, p+1, q+2, 1}  -  A_{3, p+3, q-2, 1} $\\&$+ 
    A_{3, p+3, q, 1})+ 9 (A_{3, p+1, q-2, 1}- A_{3, p-1, q-2, 1}$\\&$ + 
    A_{3, p-1, q, 1}    - 
    A_{3, p+1, q, 1}) + 
   A_{3, p+3, q-4, 1}$\\&$ + A_{3, p-3, q+2, 1}  - A_{3, p+3, q+2, 1}- 
   A_{3, p-3, q-4, 1}]$\\&$\times (\sin 3I_r-3 \sin I_r)$\\&$
  +\frac{3\alpha^2}{256}  [2 (q-2) (A_{2, p-2, q-3, 0}-  2 A_{2, p-2, q-1, 0}$\\&$+  A_{2, p-2, q+1, 0} -2 A_{2, p, q-3, 0}+ 4 A_{2, p, q-1, 0}$\\&$-2 A_{2, p, q+1, 0}+A_{2, p+2, q-3, 0}-2 A_{2, p+2, q-1, 0}$\\&$+  A_{2, p+2, q+1, 0}  ) - 
   A_{2, p-2, q-3, 1}  - 
   A_{2, p+2, q+1, 1}$\\&$- 
   A_{2, p-2, q+1, 1} - 
   A_{2, p+2, q-3, 1} + 2 (A_{2, p-2, q-1, 1}  $\\&$ + A_{2, p, q-3, 1}  + A_{2, p, q+1, 1} +  A_{2, p+2, q-1, 1})$\\&$ - 
   4 A_{2, p, q-1, 1}  ] \sin 2 I_r
  $\\
     $c_{31}^1 $&   $\frac{\alpha}{128}  [q (7 - 18 q + 8 q^2) (A_{1, p-1, q-2, 0} - A_{1, p-1, q, 0}$\nonumber\\&$ -A_{1, p+1, q-2, 0} +A_{1, p+1, q, 0} )$\nonumber\\&$ 
     + (8 - 3 q - 4 q^2) (A_{1, p-1, q-2, 1} -A_{1, p-1, q, 1}$\nonumber\\&$ -A_{1, p+1, q-2, 1}+A_{1, p+1, q, 1} )
     + (7-2q) (A_{1, p-1, q-2, 2}$\nonumber\\&$ - A_{1, p-1, q, 2} -A_{1, p+1, q-2, 2}+A_{1, p+1, q, 2})$\nonumber\\&$ 
     +  A_{1, p-1, q-2, 3}  - A_{1, p-1, q, 3}         -    A_{1, p+1, q-2, 3} $\nonumber\\&$ 
        + A_{1, p+1, q, 3}]\sin I_r$\\
 \hline
  \end{tabular}
 \end{table}

\begin{table}
  \caption{Force coefficients $c^{2}_{mn}(p,q,\alpha,I_r)$ of the term $e^ms^n \cos(\phi-2\omega)$. }\label{t3}
   \begin{tabular}{cl}
    \hline
    $c^{2}_{20} $&$\frac{1}{16}[(6 - 11 q +   4 q^2) A_{0, p, q-2, 0} + (6  -  4 q) A_{0, p, q-2, 1} $\nonumber\\&$+ A_{0, p, q-2, 2}],$\\ 
    $c_{21}^2 $&$-\frac{\alpha}{64}  [(4 q^2-15q+12) (A_{1, p-1, q-3, 0}-A_{1, p-1, q-1, 0}$\nonumber\\&$-A_{1, p+1, q-3, 0}+A_{1, p+1, q-1, 0}) 
    $\nonumber\\&$-    4 (q-2) (A_{1, p-1, q-3, 1}+A_{1, p-1, q-1, 1}$\nonumber\\&$+A_{1, p+1, q-3, 1}-A_{1, p+1, q-1, 1}) + A_{1, p-1, q-3, 2}$\nonumber\\&$ - A_{1, p-1, q-1, 2} 
    - A_{1, p+1, q-3, 2} + A_{1, p+1, q-1, 2}]\sin I_r$\\
    $c_{40}^2 $&$\frac{1}{192} [(12 + 26 q - 88 q^2 + 68 q^3 - 16 q^4) A_{0, p, q-2, 0} $\nonumber\\&$- 
   2 (6 - 23 q +24 q^2 - 8 q^3) A_{0, p, q-2, 1} $\nonumber\\&$+ 
   (6 -9q) A_{0, p, q-2, 2}  + 
   4(2-q) A_{0, p, q-2, 3}$\nonumber\\&$+ A_{0, p, q-2, 4}]$\\
     $c_{22}^2 $&$\frac{3\alpha^2}{512} [(20 - 19 q + 4 q^2) (A_{2, p-2, q-4,   0} $\nonumber\\
     &$-2A_{2, p-2, q-2, 0} +A_{2, p-2, q, 0}-2A_{2, p, q-4, 0}$\nonumber\\
     &$+4 A_{2, p, q-2, 0}-2A_{2, p, q, 0} +A_{2, p+2, q-4, 0}$\nonumber\\&$-2A_{2, p+2, q-2, 0}+A_{2, p+2, q, 0})$\nonumber\\&$
     + (10 - 4 q) (A_{2, p-2, q-4, 1}-2A_{2, p-2, q-2, 1} +A_{2, p-2, q, 1}$\nonumber\\&$-2A_{2, p, q-4, 1} +4A_{2, p, q-2, 1} -2A_{2, p, q, 1}$\nonumber\\&$+A_{2, p+2, q-4, 1}-2A_{2, p+2, q-2, 1}+A_{2, p+2, q, 1}) $\nonumber\\&$
     + A_{2, p-2, q-4, 2} 
     - 2 A_{2, p-2, q-2, 2} 
     + A_{2, p-2, q, 2}$\nonumber\\&$
     - 2 A_{2, p, q-4, 2} 
     + 4 A_{2, p, q-2, 2} 
     -  2 A_{2, p, q, 2} $\nonumber\\&$
      +A_{2, p+2, q-4, 2}
     -2 A_{2, p+2, q-2, 2} 
  + A_{2, p+2, q, 2}]\sin^2I_r$\\&
 $-\frac{\alpha}{128} [(12 - 15 q + 4 q^2) (A_{1, p-1, q-3, 0} $\\&$- A_{1, p+1, q-3, 0}-  A_{1, p-1, q-1, 0} + 
    A_{1, p+1, q-1, 0})$\\&$ - 
   4 (q-2) (A_{1, p-1, q-3, 1}+A_{1, p-1, q-1, 1}+A_{1, p+1, q-3, 1}$\\&$-A_{1, p+1, q-1, 1} ) + A_{1, p-1, q-3, 2}  - A_{1, p-1, q-1, 2}$\\&$- A_{1, p+1, q-3, 2}  + A_{1, p+1, q-1, 2}] \cos I_r$\\
 \hline
  \end{tabular}
 \end{table}

\begin{table}
  \caption{Force coefficients $c^{3}_{mn}(p,q,\alpha,I_r)$ of the term $e^ms^n \cos(\phi-3\omega)$. }\label{t4}
   \begin{tabular}{cl}
    \hline
    $c^{3}_{30} $&$\frac{1}{96} [(8 q^3- 42 q^2+ 62 q -24 ) A_{0, p, q-3, 0} $\nonumber\\&$- 
   3 (12 - 15 q + 4 q^2) A_{0, p, q-3, 1} +(6q- 12) A_{0, p, q-3, 2} $\nonumber\\&$ - A_{0, p, q-3, 3}]$\\ 
   $c_{31}^3 $&   -$\frac{\alpha}{384} [(60 - 107 q + 54 q^2 - 8 q^3) (A_{1, p-1, q-2, 0} -A_{1, p-1, q-4, 0}$\nonumber\\&$+A_{1, p+1, q-4, 0}-A_{1, p+1, q-2, 0}   )$\nonumber\\&$
   - 3 (20 - 19 q + 4 q^2) (A_{1, p-1, q-4, 1}- A_{1, p-1, q-2, 1}$\nonumber\\&$ -A_{1, p+1, q-4, 1}+A_{1, p+1, q-2, 1} ) 
   + (15-6q) (A_{1, p-1, q-2, 2} $\nonumber\\&$-A_{1, p-1, q-4, 2}+A_{1, p+1, q-4, 2} -A_{1, p+1, q-2, 2})$\nonumber\\&$
   - A_{1, p-1, q-4, 3} 
   + A_{1, p-1, q-2, 3} 
   + A_{1, p+1, q-4, 3} $\nonumber\\&$
   -  A_{1, p+1, q-2, 3}]\sin I_r$\\
 \hline
  \end{tabular}
 \end{table}

\begin{table}
  \caption{Force coefficients $c^{4}_{mn}(p,q,\alpha,I_r)$ of the term $e^m\cos^n I \cos(\phi-4\omega)$. }\label{t5}
   \begin{tabular}{cl}
    \hline
    $c^{4}_{40} $&$\frac{1}{768} [(120 - 394 q + 379 q^2 - 136 q^3 + 16 q^4) A_{0, p, q-4, 
     0} $\nonumber\\&$ +4 (60 - 107 q + 54 q^2 - 8 q^3) A_{0, p, q-4, 1} $\nonumber\\&$+ 
   (120-114q+24q^2) A_{0, p, q-4, 2}$\nonumber\\&$+ (20-8q) A_{0, p, q-4, 3}  + A_{0, p, q-4, 4}]$\\ 
 \hline
  \end{tabular}
 \end{table}

\begin{table}
  \caption{Force amplitudes and cosine arguments of  the indirect part. }\label{t6}
   \begin{tabular}{ll}
    \hline
     Cosine argument & Force amplitude\\ \hline\\
$ \lambda^\prime-\varpi$&$-\frac{3\alpha}{8} [2 s \sin I_r+(s^2-2)\cos I_r-2] e $\\ 

$\lambda - \lambda^\prime$&$\frac{\alpha}{128} [
  e^4 + 32 e^2-64 + [e^4 - 16 e^2 ( s^2-2) $\\&$+ 8 (s^4 + 4 s^2 -8 )] \cos I_r$\\&$ - 
  32 (e^2-2) s \sin I_r)]$\\   

$\lambda - \lambda^\prime- 2 \varpi + 2 \Omega$&$\frac{\alpha}{96} e^2 [(6 + 2 e^2 - 3 s^2) \cos I_r
    $\\&$- 
   2 (3 + e^2 + 3 s \sin I_r)]$\\
    
$2 \lambda - \lambda^\prime- \varpi $&$  \frac{\alpha}{16} e  [( 3 e^2 + 2 s^2-4) \cos I_r $\\&$- (3 e^2-4) ( s \sin I_r-1)]$\\ 
   
$2 \lambda - \lambda^\prime- 3 \varpi + 2 \Omega $&$\frac{\alpha}{48} e^3 ( \cos I_r - s \sin I_r-1)$\\
       
$3 \lambda - \lambda^\prime- 2 \varpi $&$\frac{3\alpha}{32} e^2  [( + 2 e^2 + s^2-2) \cos I_r $\\&$+ 
   2 ( e^2 + s \sin I_r-1)]  $\\
      
$3 \lambda - \lambda^\prime- 4 \varpi + 2 \Omega $&$ -\frac{3\alpha}{128} e^4  \sin^2 (I_r/2) $\\
          
$4 \lambda - \lambda^\prime- 3 \varpi $&$ -\frac{\alpha}{6} e^3  (1 + \cos I_r - s \sin I_r) $\\
     
$5 \lambda - \lambda^\prime- 4 \varpi $&$ -\frac{125\alpha}{384} e^4  \cos^2 (I_r/2) $\\
     
$ \lambda^\prime+\varpi - 2 \Omega $&$ \frac{3\alpha}{8} e  [2 + ( s^2-2) \cos I_r + 2 s \sin I_r] $\\
     
$\lambda +\lambda^\prime- 2 \Omega $&$ -\frac{\alpha}{128}  [64 - 32 e^2 - 
   e^4 + $\\&$[e^4 - 16 e^2 (s^2-2)$\\&$ + 8 ( 4 s^2 + s^4-8)] \cos I_r $\\&$- 
   32 (e^2-2) s \sin I_r] $\\
      
$\lambda + \lambda^\prime- 2 \varpi $&$ -\frac{\alpha}{96} e^2  [(6 + 2 e^2 - 3 s^2) \cos I_r$\\&$ + 
   2 (3 + e^2 - 3 s \sin I_r)] $\\
    
$2 \lambda + \lambda^\prime- 3 \varpi $&$ -\frac{\alpha}{48} e^3 (1 + \cos I_r - s \sin I_r) $\\
     
$2 \lambda + \lambda^\prime- \varpi - 2 \Omega$&$ \frac{\alpha}{16} e  [(4 - 3 e^2 - 2 s^2) \cos I_r$\\&$+ ( 3 e^2-4) (1 + s \sin I_r)] $\\
    
$3 \lambda + \lambda^\prime- 4 \varpi $&$ -\frac{3\alpha}{128} e^4 \alpha \cos^2 (I_r/2) $\\
        
$3 \lambda + \lambda^\prime- 2 \varpi -2 \Omega $&$-\frac{3\alpha}{32} e^2  [2 - 2 e^2 + ( 2 e^2 + s^2-2) \cos I_r $\\&$+ 
   2 s \sin I_r]  $\\
  
$4 \lambda + \lambda^\prime- 3 \varpi - 2 \Omega $&$ \frac{\alpha}{6} e^3  (\cos I_r - s \sin I_r-1) $\\

$5 \lambda + \lambda^\prime- 4 \varpi - 2 \Omega $&$  -\frac{125\alpha}{384} e^4 \alpha \sin^2 (I_r/2)$\\ 
 \hline
  \end{tabular}
 \end{table}

\begin{table}
  \caption{Force coefficients $s^{k}_{mn}(\alpha,I_r)$ of the secular term $e^{m}s^n\cos (k\omega)$. }\label{t7}
   \begin{tabular}{ccl}
    \hline
   $k=0$  & $s^{0}_{01} $ &$\frac{ \alpha}{8} (A_{1, 1, -1, 0} - A_{1, 1, 1, 0}) \sin I_r$\\ 
              & $s^{0}_{20} $ &$\frac{1}{16}(2 A_{0, 0, 0, 1} + A_{0, 0, 0, 2})$\\ 
               & $s^{0}_{02} $ &$\frac{\alpha}{64} [4 (A_{1, 1, -1, 0} - A_{1, 1, 1, 0}) \cos I_r + $\nonumber \\ &&$+ 
   3 \alpha (2 A_{2, 0, 0, 0} + A_{2, 2, -2, 0} - 4 A_{2, 2, 0, 0} +$\nonumber \\&&$+ A_{2, 2, 2, 0}) \sin^2 I_r] $   \nonumber 
\\ 
& $s^{0}_{03} $ &
$\frac{\alpha^2}{256}  \sin I_r [12 (2 A_{2, 0, 0, 0} + A_{2, 2, -2, 0} $\\
 & &$- 4 A_{2, 2, 0, 0} + 
      A_{2, 2, 2, 0}) \cos I_r + 
   5 \alpha (9 A_{3, 1, -1, 0} $\\&&$- 9 A_{3, 1, 1, 0} + A_{3, 3, -3, 0} - 
      6 A_{3, 3, -1, 0}$\\&&$ + 6 A_{3, 3, 1, 0} - A_{3, 3, 3, 0}) \sin^2I_r]$\\
          &$s^{0}_{21} $&$\frac{\alpha}{32}  (2 A_{1, 1, -1, 0} + 4 A_{1, 1, -1, 1} + A_{1, 1, -1, 2}$\\&&$ - 
   2 A_{1, 1, 1, 0} - 4 A_{1, 1, 1, 1} - A_{1, 1, 1, 2}) \sin I_r$\\
          & $s^{0}_{22} $&$
       \frac{\alpha}{512)}  [3 \alpha (12 A_{2, 0, 0, 0} + 12 A_{2, 0, 0, 1}$\\&&$ + 
      2 A_{2, 0, 0, 2} + 6 A_{2, 2, -2, 0} + 6 A_{2, 2, -2, 1} $\\&&$+ 
      A_{2, 2, -2, 2} - 24 A_{2, 2, 0, 0} - 24 A_{2, 2, 0, 1}$\\&&$ - 
      4 A_{2, 2, 0, 2} + 6 A_{2, 2, 2, 0} + 6 A_{2, 2, 2, 1} $\\&&$+ 
      A_{2, 2, 2, 2}) - 
   3 \alpha (12 A_{2, 0, 0, 0} + 12 A_{2, 0, 0, 1} $\\&&$+ 
      2 A_{2, 0, 0, 2} + 6 A_{2, 2, -2, 0} + 6 A_{2, 2, -2, 1}$\\&&$ + 
      A_{2, 2, -2, 2} - 24 A_{2, 2, 0, 0} - 24 A_{2, 2, 0, 1} $\\&&$- 
      4 A_{2, 2, 0, 2} + 6 A_{2, 2, 2, 0} + 6 A_{2, 2, 2, 1} $\\&&$+ 
      A_{2, 2, 2, 2}) \cos (2I_r) +
   8 (2 A_{1, 1, -1, 0} + 4 A_{1, 1, -1, 1} $\\&&$+ A_{1, 1, -1, 2} - 
      2 A_{1, 1, 1, 0} - 4 A_{1, 1, 1, 1} $\\&&$- A_{1, 1, 1, 2}) \cos I_r]    
          $\\ 
          &$s^{0}_{40} $&$\frac{1}{256} (4 A_{0, 0, 0, 3} + A_{0, 0, 0, 4})$\\ 
          &$s^{0}_{04} $&$
          \frac{3\alpha^2}{32768}  [64(2A_{2, 0, 0, 0} +  A_{2, 2, -2, 0}$\\&&$ - 
   4 A_{2, 2, 0, 0} +  A_{2, 2, 2, 0}) $\\&&$+ 7\alpha^2[
   90  A_{4, 0, 0, 0} + 40 ( 2A_{4, 2, -2, 0}$\\&&$+   2A_{4, 2, 2, 0}- A_{4, 4, -2, 0}  - A_{4, 4, 2, 0}) $\\&&$- 
   60 (4A_{4, 2, 0, 0}    +   A_{4, 4, 0, 0})+ 5 (A_{4, 4, 4, 0}$\\&&$+  A_{4, 4, -4, 0})]]
   +\frac{\alpha}{2048} [32 (A_{1, 1, -1, 0} -  A_{1, 1, 1, 0})$\\&&$+ 
   15 \alpha^2 (9 A_{3, 1, -1, 0} - 9 A_{3, 1, 1, 0}$\\&&$ - 6 A_{3, 3, -1, 0} + 6 A_{3, 3, 1, 0} - 
      A_{3, 3, 3, 0}$\\&&$+    A_{3, 3, -3, 0} )] \cos I_r
  -\frac{\alpha^2}{8192}[
    48(4A_{2, 2, 0, 0}-2A_{2, 0, 0, 0}$\\&&$-A_{2, 2, -2, 0}-A_{2, 2, 2, 0} )
    + 35\alpha^2[
     A_{4, 4, -4, 0} $\\&&$ + A_{4, 4, 4, 0}- 8 A_{4, 4, -2, 0} - 8 A_{4, 4, 2, 0}$\\&&$+ 
    12  A_{4, 4, 0, 0}+ 16  A_{4, 2, -2, 0} + 16 A_{4, 2, 2, 0}  $\\&&$
    +18 A_{4, 0, 0, 0}  -  48 A_{4, 2, 0, 0}  ]] \cos 2I_r$\\&&$
  -\frac{15\alpha^3}{2048} (9 A_{3, 1, -1, 0} - 9 A_{3, 1, 1, 0}$\\&&$ + 
    A_{3, 3, -3, 0} - 6 A_{3, 3, -1, 0} + 6 A_{3, 3, 1, 0}$\\&&$ - 
    A_{3, 3, 3, 0}) \cos 3I_r 
    +\frac{35\alpha^4}{32768}  ( A_{4, 4, 4, 0}+ A_{4, 4, -4, 0}$\\&&$ - 
   8 A_{4, 4, -2, 0}  - 8 A_{4, 4, 2, 0} + 12 A_{4, 4, 0, 0}$\\&&$ + 
   16 A_{4, 2, -2, 0}+ 16 A_{4, 2, 2, 0}+18 A_{4, 0, 0, 0}$\\&&$ - 
   48 A_{4, 2, 0, 0}) \cos 4I_r
          
          $\\

$k=2$&$s^{2}_{20} $&$\frac{1}{16}(6 A_{0, 0, 2, 0} + 6 A_{0, 0, 2, 1} + A_{0, 0, 2, 2})$\\
    &$s^{2}_{21} $&$-\frac{\alpha}{64}  [12 (A_{1, 1, -1, 0} -  A_{1, 1, 1, 0}$\\&&$
    -  A_{1, 3, -1, 0}+ A_{1, 3, 1, 0})+ 8( A_{1, 1, -1, 1}$\\&&$
     -  A_{1, 1, 1, 1} -  A_{1, 3, -1, 1}+  A_{1, 3, 1, 1})$\\&&$+ 
   A_{1, 1, -1, 2}  - 
   A_{1, 1, 1, 2}   - 
   A_{1, 3, -1, 2} $\\&&$  + 
   A_{1, 3, 1, 2}] \sin I_r$\\

     &$s^{2}_{22} $&$
     
     -\frac{\alpha}{512} [3 \alpha (10 [ 4 A_{2, 2, 2, 0} + 2 A_{2, 2, 2, 1} $\\&&$ +4A_{2, 0, 0, 0} + 2 A_{2, 0, 0, 1} + 4 A_{2, 2, -2, 0}  $\\&&$+ 2 A_{2, 2, -2, 1} + 4 A_{2, 4, 0, 0} + 2 A_{2, 4, 0, 1} $\\&&$ - 2 A_{2, 4, -2, 0} -A_{2, 4, -2, 1} - 2 A_{2, 4, 2, 0} $\\&&$ -  A_{2, 4, 2, 1}- 12 A_{2, 2, 0, 0} - 6 A_{2, 2, 0, 1}] $\\&&$- 
      6 A_{2, 2, 0, 2} + 
      2 [A_{2, 0, 0, 2} + 
       A_{2, 2, -2, 2}   $\\&&$+ 
       A_{2, 2, 2, 2}  + 
       A_{2, 4, 0, 2}] - 
      A_{2, 4, 2, 2}  $\\&&$- 
      A_{2, 4, -2, 2})\sin^2I_r +
   4 (12 [A_{1, 1, -1, 0} $\\&&$- 
       A_{1, 1, 1, 0} - 
       A_{1, 3, -1, 0}+ 
       A_{1, 3, 1, 0}]  $\\
   \hline
  \end{tabular}
 \end{table}    
  
     \setcounter{table}{6}
     
\begin{table}
  \caption{Continued. }
   \begin{tabular}{ccl}
    \hline\\    
           &&$+ 8 [A_{1, 1, -1, 1}  -  A_{1, 1, 1, 1} +  A_{1, 3, 1, 1} $\\&&$-  A_{1, 3, -1, 1}]+ A_{1, 1, -1, 2} - A_{1, 1, 1, 2} $\\&&$ - A_{1, 3, -1, 2}   + A_{1, 3, 1, 2}) \cos I_r]    $\\
     &$s^{2}_{40} $&$\frac{1}{192} (12 A_{0, 0, 2, 0} - 12 A_{0, 0, 2, 1} + 6 A_{0, 0, 2, 2} $\nonumber\\&&$+ 
   8 A_{0, 0, 2, 3} + A_{0, 0, 2, 4})$\\ 
   
    $k=4$&$s^{4}_{40}  $&$\frac{1}{768} (120 A_{0, 0, 4, 0} + 240 A_{0, 0, 4, 1}$\nonumber\\&&$ + 120 A_{0, 0, 4, 2} + 
   20 A_{0, 0, 4, 3} + A_{0, 0, 4, 4})$\\
 \hline
  \end{tabular}
 \end{table}

\section*{Acknowledgments}
The authors thank Zoran Kne{\v{z}}evi\'c for his thorough and helpful review of the manuscript. M.H.M.M. was supported by grant 2015/17962-5 of S\~ao Paulo Research Foundation (FAPESP). 

\bibliographystyle{mnras}

\bibliography{ms}

\end{document}